\documentclass{article}

\usepackage[preprint]{neurips_2026}

\usepackage[T1]{fontenc}

\usepackage[utf8]{inputenc}

\usepackage{microtype}      
\usepackage{hyperref}       
\usepackage{url}            
\usepackage{booktabs}       
\usepackage{amsfonts}       
\usepackage{nicefrac}       
\usepackage{xcolor}         
\usepackage{enumitem}

\usepackage{graphicx}
\usepackage{pgfplots}
\pgfplotsset{compat=1.18}
\usepackage{amsmath}
\usepackage{pifont}         
\usepackage{makecell}
\usepackage{tabularx}
\usepackage{array}
\usepackage{colortbl}
\usepackage{fancyvrb}
\definecolor{diffaddbg}{HTML}{e6ffec}
\definecolor{diffdelbg}{HTML}{ffebe9}
\definecolor{diffhunkbg}{HTML}{eaeef2}
\definecolor{diffadd}{HTML}{1a7f37}
\definecolor{diffdel}{HTML}{cf222e}
\definecolor{diffhunk}{HTML}{0550ae}
\newcommand{\DA}[1]{{\setlength{\fboxsep}{0pt}%
  \colorbox{diffaddbg}{\makebox[\linewidth][l]{\strut\scriptsize\ttfamily\textcolor{diffadd}{+}#1}}}}
\newcommand{\DD}[1]{{\setlength{\fboxsep}{0pt}%
  \colorbox{diffdelbg}{\makebox[\linewidth][l]{\strut\scriptsize\ttfamily\textcolor{diffdel}{-}#1}}}}
\newcommand{\DC}[1]{\strut\scriptsize\ttfamily#1}
\renewcommand{\DH}[1]{{\setlength{\fboxsep}{0pt}%
\colorbox{diffhunkbg}{\makebox[\linewidth][l]{\strut\scriptsize\ttfamily\textcolor{diffhunk}{\bfseries#1}}}}}

\newcommand{\myNum}[1]{\emph{(#1)}}
\newcommand{\benchmark}{MobileDev-Bench}

\author{
  Moshood Fakorede$^{1}$ \quad
  Krishna Upadhyay$^{1}$ \quad
  A.B. Siddique$^{2}$ \quad
  Umar Farooq$^{1}$ \\[0.5em]
  $^{1}$Louisiana State University, $^{2}$University of Kentucky \\[0.3em]
  \texttt{\{mfakor1, kupadh4, ufarooq\}@lsu.edu} \quad
  \texttt{siddique@cs.uky.edu}
}

\title{MobileDev-Bench: A Benchmark for Issue Resolution in Mobile Application Development}

\begin{document}

\maketitle

\begin{abstract}

Large language models (LLMs) have shown strong performance on automated software engineering tasks, yet existing benchmarks focus primarily on library-style repositories, leaving mobile application development largely unexplored despite its framework-specific build systems, heterogeneous artifact types, and coordinated multi-file fix requirements. 
We introduce {\benchmark}, a benchmark comprising 407 real-world issue-resolution tasks collected from 19 production mobile applications spanning Android Native (Java/Kotlin), React Native (TypeScript), and Flutter (Dart). 
Each task pairs a verified developer-reported issue with executable test patches, enabling fully automated validation of model-generated fixes within mobile build environments. 
The benchmark exhibits substantially greater patch complexity than prior benchmarks: fixes modify 12.9 files and 334.6 lines on average, and 41\% of instances require coordinated changes across multiple artifact types, such as source, build configuration, and resource files.
Evaluation of four frontier LLMs (Claude Sonnet 4.5, Qwen3-Coder, GPT-5.2, and Gemini 2.5 Flash) yields end-to-end resolution rates of only 3.23\%–4.23\% under automated retrieval and at most 5.69\% under oracle retrieval, well below resolution rates reported on existing benchmarks.
We release {\benchmark} with task instances, an evaluation harness, and containerized environments to support reproducible research on AI-assisted mobile application development.

\end{abstract}

\section{Introduction}
\label{sec:introduction}


Large language models (LLMs) such as GPT-4~\citep{openai2023gpt4}, Claude~\citep{anthropic2023claude}, and Code Llama~\citep{roziere2023code} have demonstrated strong progress in automated software engineering, particularly in repository-level issue resolution. 
Benchmarks such as SWE-bench~\citep{jimenez2024swebench} establish a realistic evaluation paradigm by requiring models to resolve real-world GitHub issues through patch generation, where correctness is validated by executing the relevant test suite.
Subsequent extensions~\citep{yang2024swebenchmultimodal, zhang2024codev, kabir2025swebenchmultilingual, mhatre2025swesharpbench, 2024swebenchjava, multi2025swebench} broaden coverage across programming languages and evaluation settings, revealing performance gaps even for frontier code-capable LLMs.


Existing benchmarks primarily target issue resolution in general-purpose libraries or web applications.
Mobile application~(app) development remains largely unexplored despite several properties that distinguish it from library-style repositories. 
Framework-specific build systems govern compilation and testing, fixes span heterogeneous artifact types including source code, manifests, and resource definitions, and correctness often requires coordinating changes across multiple files within a single patch~\citep{android-resources, android-build, flutter-state-declarative}. 
Moreover, they frequently employ declarative and event-driven programming~\citep{flutter-declarative} and rely on asynchronous state management across user interface and service components.
These characteristics make mobile app repositories structurally different from the library-centric settings that dominate existing issue-resolution benchmarks.

\begin{table*}[t!]
\centering
\footnotesize
\resizebox{\textwidth}{!}{
\begin{tabular}{
l
p{2.25cm}
>{\centering\arraybackslash}p{0.7cm}
>{\centering\arraybackslash}p{0.7cm}
>{\centering\arraybackslash}p{0.7cm}
>{\centering\arraybackslash}p{0.7cm}
>{\centering\arraybackslash}p{1.47cm}
>{\raggedleft\arraybackslash}p{0.9cm}
>{\raggedleft\arraybackslash}p{0.9cm}
}
\toprule
\textbf{Benchmarks} &
\textbf{Domain} &
\textbf{\#Tasks} &
\textbf{\#Repos} &
\textbf{Verified} &
\textbf{Difficulty level} &
\textbf{Multi-type artifact fixes} &
\textbf{Patch \#~Files} &
\textbf{Complex. \# Lines} \\
\midrule
SWE-bench  &
Python libraries &
2294 &
12 &
\ding{55} &
\ding{55} &
\ding{55} &
1.7 & 32.8 \\

SWE-bench Verified &
Python libraries &
500 &
12 &
\ding{51} &
\ding{55} &
\ding{55} &
1.2 & 14.3 \\

SWE-bench Multimodal &
JavaScript libraries &
617 &
17 &
\ding{51} &
\ding{55} &
\ding{51} &
3.9 & 131.6 \\

SWE-PolyBench &
Multi-lang. libraries &
2110 &
21 &
\ding{55} &
\ding{55} &
\ding{55} &
2.6 & 51.2 \\

SWE-bench Multilingual &
Multi-lang. libraries &
300 &
42 &
\ding{55} &
\ding{55} &
\ding{55} &
1.7 & 47.8 \\

OmniGIRL &
Multi-lang. repos &
959 &
15 &
\ding{55} &
\ding{55} &
\ding{55} &
1.2 & 46.3 \\

SWE-bench-java &
Java libraries &
91 &
6 &
\ding{51} &
\ding{55} &
\ding{55} &
2.3 & 38.7 \\

SWE-Sharp-bench&
C\# libraries &
150 &
17 &
\ding{51} &
\ding{55} &
\ding{55} &
4.9 & 131.2 \\

Rust-SWE-bench &
Rust repositories &
500 &
34 &
\ding{51} &
\ding{55} &
\ding{55} &
9.8 & 139.9 \\

Multi-SWE-bench &
Multi-lang. repos &
1632 &
39 &
\ding{51} &
\ding{51} &
\ding{55} &
4.9 & 163.3 \\

\midrule
\textbf{\benchmark} &
\textbf{Mobile Apps} &
\textbf{407} &
\textbf{19} &
\textbf{\ding{51}} &
\textbf{\ding{51}} &
\textbf{\ding{51}} &
\textbf{12.9} & \textbf{334.6} \\
\bottomrule
\end{tabular}
}
\caption{
Comparison of issue-resolution benchmarks.
Unlike prior work that primarily evaluates source-code fixes in library-style repositories, {\benchmark} targets mobile apps, incorporates human verification and difficulty stratification, evaluates multi-artifact fixes (e.g., source and resources artifacts such as manifest files), and restores pre-fix buildability to enable execution-based evaluation.}
\label{tab:benchmark-comparison}
\vspace{-10pt}
\end{table*}


We introduce \emph{\benchmark}, a benchmark of 407 manually verified issue-resolution tasks collected from 19 production mobile app repositories spanning Android Native (Java/Kotlin), React Native (TypeScript), and Flutter (Dart). 
Each instance pairs a verified developer-reported issue with its merged pull request and executable test patches, enabling automated validation of model-generated fixes.
To support reliable evaluation, {\benchmark} executes tests under three configurations: the base commit, the test patch only, and the full fix, capturing transitions such as Fail-to-Pass, None-to-Pass, Pass-to-Pass, and Pass-to-Fail. 
Unlike prior benchmarks that exclude instances with compilation failures~\citep{wang2025swebenchframeworkscalablegeneration}, we explicitly model transitions from non-existent or non-executable states and restore projects to a compilable state, enabling execution-based evaluation.
This compilation-aware design is particularly important for \emph{statically typed programming languages} common in mobile app development, such as Java and Kotlin, where successful compilation is required before tests can execute.


Table~\ref{tab:benchmark-comparison} provides a comparison of {\benchmark} with prior issue-resolution benchmarks.
It targets production mobile apps drawn from mature and actively maintained open-source repositories, each with at least 400 GitHub stars.
It evaluates fixes that span multiple artifact types, including source code, resources, and configuration metadata.
It incorporates \emph{manual verification} and difficulty stratification to ensure benchmark quality, with tasks distributed across easy (138), medium (139), and hard (130) tiers.
Additionally, it exhibits substantially higher patch complexity: fixes modify 12.9 files and 334.6 lines on average, and 41\% of instances require coordinated changes across multiple artifact types.
These properties capture the multi-file and multi-artifact dependencies common in mobile app development that are largely absent from existing library-centric benchmarks.


We extend Agentless~\citep{xia2024agentless} to support Java, Kotlin, TypeScript, and Dart and evaluate four frontier LLMs on {\benchmark}. 
Despite strong performance on prior issue-resolution benchmarks, models achieve low resolution rates of 3.23\%--4.23\% under automated retrieval and at most 5.69\% under oracle retrieval.
Both file-level localization and patch generation contribute to this low performance. 
Under automated retrieval, recall ranges from 13.8\% to 18.6\% and declines steeply as the number of files to modify grows, limiting achievable resolution rates.
Under oracle retrieval, recall rises to 31.4\%–55.4\% overall and 84.5\%–100\% on single-file tasks, yet resolution still reaches at most 5.69\%, indicating that patch generation is the larger remaining bottleneck and that correctly identifying files to modify is necessary but not sufficient for successful resolution.



This work makes the following contributions:

\begin{itemize}
\setlength{\itemsep}{1pt}
\setlength{\parskip}{0pt}
\setlength{\parsep}{0pt}
\setlength{\topsep}{1pt}
\item We introduce {\benchmark}, a benchmark of 407 verified mobile app issue-resolution tasks drawn from 19 production repositories across Android, React Native, and Flutter.
\item We develop a compilation-aware, execution-based evaluation framework for issue resolution across Java, Kotlin, Dart, and TypeScript mobile app projects and release {\benchmark} with task instances, an evaluation harness, and containerized environments at \url{https://huggingface.co/datasets/MobileDev-Bench/mobiledev-bench}.
\item We provide a comprehensive evaluation of four frontier code-capable LLMs, showing resolution rates of 3.23\%–4.23\% under automated and at most 5.69\% under oracle retrieval.
\end{itemize}

\section{Related Work}
\label{sec:related-work}

\begin{figure*}[t]
\centering
\includegraphics[width=0.98\textwidth]{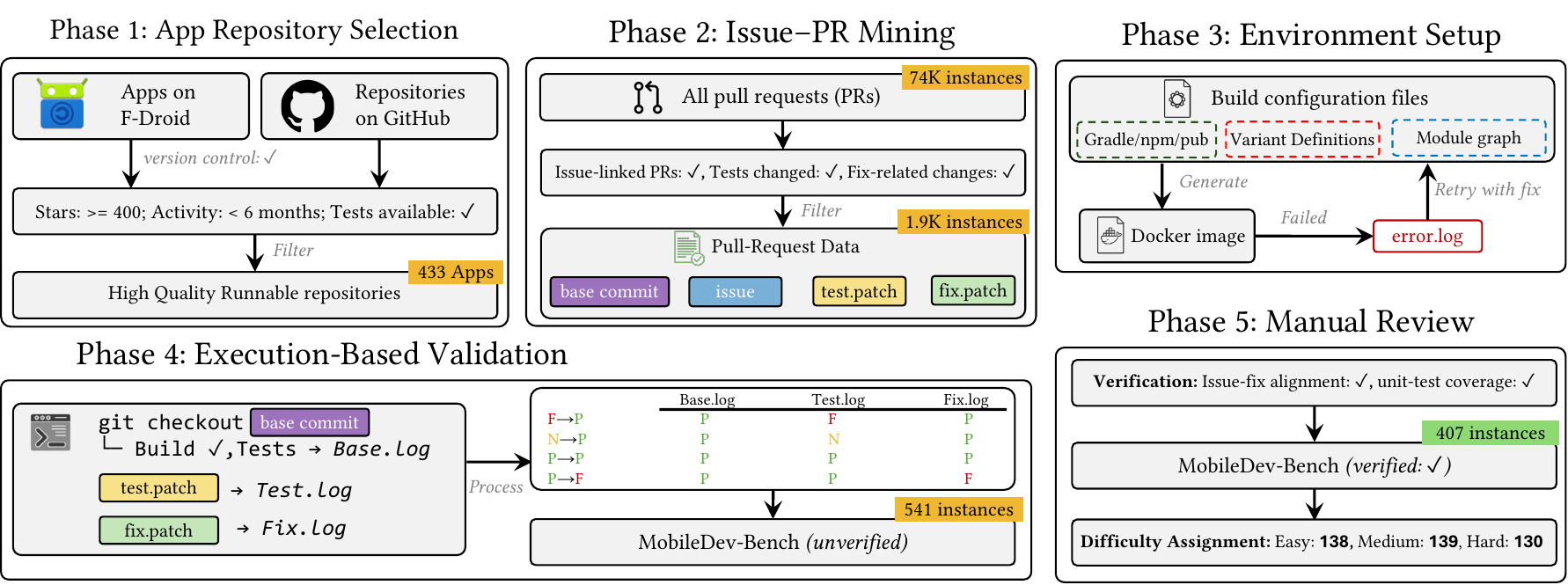}
\caption{Overview of {\benchmark} construction pipeline. The framework consists of five phases: (1) filtering for high-quality repositories, (2) mining issue-linked PRs, (3) reproducing mobile builds via Docker, (4) executing compilation-aware validation, and (5) manually reviewing data to yield 407 verified, difficulty-stratified benchmark instances.}
\label{fig:pipeline}
\end{figure*}


Early benchmarks such as HumanEval~\citep{chen2021evaluating} and MBPP~\citep{austin2021program} established a foundation for assessing the programming capabilities of LLMs. 
These benchmarks primarily focused on small, self-contained problems: 95.8\% target Python and over 83\% evaluate function- or statement-level tasks~\citep{cao2024javabench}, overlooking the real-world repository-level complexity.

SWE-bench~\citep{jimenez2024swebench} introduced repository-scale evaluation with 2,294 instances derived from real GitHub issues in Python projects, requiring models to generate patches validated through test execution. 
Subsequent work extended this direction to additional languages. 
Multi-SWE-bench~\citep{multi2025swebench} spans seven programming languages with 1,632 instances, while SWE-bench~Java~\citep{2024swebenchjava} provides 91 Java-specific tasks. 
SWE-PolyBench~\citep{2025swepolybench} further shows that realistic repair tasks often require edits across multiple files. 
Despite this progress, existing benchmarks primarily focus on library-style repositories~\citep{2025omnigirl, mhatre2025swesharpbench} and do not capture platform-specific app development environments.
Alongside these benchmarks, LLM-based repair systems including SWE-agent~\citep{yang2024sweagent}, AutoCodeRover~\citep{zhang2024autocoderover}, MAGIS~\citep{tao2024magis}, OpenDevin~\citep{wang2024opendevin}, CodeR~\citep{chen2024coder}, MASAI~\citep{arora2024masai}, and Agentless~\citep{xia2024agentless} have demonstrated strong results on these benchmarks, yet none target mobile application repositories.

Mobile app development \emph{introduces additional challenges} not represented in current benchmarks, including framework-specific constraints, complex build systems, and coordination across heterogeneous artifacts such as source code, resources, and configuration files. 
Mobile app development spans Android, iOS, and cross-platform frameworks such as React Native~\citep{facebook2015reactnative} and Flutter~\citep{google2018flutter}.
Prior research in mobile apps has focused on static analysis and testing tools such as FlowDroid~\citep{arzt2014flowdroid}, AutoComply~\citep{fakorede2025autocomply}, and DroidBot~\citep{li2017droidbot}.
More recently, LLM-based approaches have targeted mobile UI task automation~\citep{wen2024autodroid} and mobile agent evaluation~\citep{deng2024mobilebench}, while benchmarks for evaluating AI-assisted \emph{code issue resolution} in mobile repositories remain \emph{largely unexplored}.

{\benchmark} addresses this gap by targeting production mobile app repositories with heterogeneous build environments and multi-artifact fixes, complementing existing library-centric benchmarks and broadening evaluation for AI-assisted software engineering systems. 
A detailed comparison of {\benchmark} with existing issue-resolution benchmarks, including patch complexity analysis and per-repository statistics, is provided in Appendix~\ref{sec:benchmarks-comparison}.

\section{{\benchmark} Construction}
\label{sec:benchmark}

{\benchmark} is constructed from real issue-pull request~(PR) pairs in open-source mobile app repositories. 
Figure~\ref{fig:pipeline} presents an overview of the five-phase construction pipeline.


\subsection{Phase 1: App Repository Selection}

We begin by constructing a pool of candidate mobile app repositories spanning three major frameworks: Android Native (Java/Kotlin), React-Native (TypeScript/JavaScript), and Flutter (Dart). 
Repositories are collected from F-Droid~\citep{fdroid}, a catalog of free and open-source Android apps, and from GitHub searches filtered by mobile framework tags.
We prioritize production apps distributed through app stores, as they typically represent mature codebases with active development and testing practices. 
This process yields an initial pool of 433 candidate repositories.

To improve quality and benchmark viability, we filter repositories based on community adoption ($\geq$400 GitHub stars), recent maintenance, an open-source license, a GitHub-hosted issue tracker, and verified buildability in a clean environment.
The detailed filter criteria are provided in Appendix~\ref{sec:construction-details}.



\subsection{Phase 2: PR Collection and Filtering}
From the selected repositories, we retrieve merged pull requests~(PR) using the GitHub API, collecting metadata such as issue descriptions, pull request text, commit history, and patch information.
To ensure stable repository states, we only consider PRs merged into the repository's default branch.

To construct valid issue-resolution tasks, we apply the following filtering steps:
\myNum{i}~We retain only PRs that are explicitly linked to a GitHub issue using standard closure patterns (e.g., ``fixes \#123'', ``resolves \#45''). 
This ensures that each task has a clear problem description derived from the issue report.
\myNum{ii}~PRs must modify or introduce test files so that the correctness of the fix can be verified through automated execution. 
\myNum{iii}~For each retained PR, we extract two patches: a test patch containing the test changes and a fix patch containing the changes. These patches are anchored to the PR's base commit to enable controlled replay of the development history.
Thus, each candidate instance contains four components: (a)~the base commit, (b)~the issue description, (c)~the test patch, and (d)~the fix patch. 
These filters yield a set of candidate issue–PR pairs for subsequent validation.
\subsection{Phase 3: Environment Runtime Setup}

Mobile app builds require precise environment configurations spanning language runtimes, framework SDKs, build tools, and dependencies. To support reproducibility, we automatically construct build environments by analyzing repository metadata and generating framework-specific Dockerfiles.
For each candidate instance, we inspect the repository at the base commit and parse framework configuration files (\texttt{build.gradle}/\texttt{.kts} and \texttt{gradle/libs.versions.toml} for Android; \texttt{package.json} for React Native; \texttt{pubspec.yaml} for Flutter) to infer toolchains and dependencies. 
When configuration files are incomplete, we supplement them with README documentation and GitHub Actions workflow files. 
We then generate a per-repository Dockerfile. 
The environment remediation details are provided in Appendix~\ref{sec:construction-details}.



\subsection{Phase 4: Execution-Based Validation}
\label{sec:execution-Based-validation}
We validate candidate instances through execution-based filtering by running tests under three configurations within isolated Docker environments: \myNum{i}~the base commit, \myNum{ii}~the base commit with the test patch applied (\texttt{test.patch}), and \myNum{iii}~the base commit with both test and fix patches applied (\texttt{fix.patch}). Rather than generic framework commands, we run instance-specific test commands derived from the PR's CI configuration, targeting only the tests relevant to each instance. 
Android instrumentation tests that require emulators are excluded due to reproducibility concerns.

For each execution, we record four states: \texttt{PASS}, \texttt{FAIL}, \texttt{NONE} (test absent in that configuration), and \texttt{SKIP} (see Appendix~\ref{sec:construction-details} for more details). We derive test-state transitions and retain instances satisfying three criteria: \myNum{i}~at least one test exhibits a \emph{Fail-to-Pass (F2P)} or \emph{None-to-Pass (N2P)} transition indicating the fix resolves the issue, \myNum{ii}~ no test exhibits a \emph{Pass-to-Fail (P2F)} regression, and \myNum{iii}~both the base and fix configurations build and execute tests successfully (we provide definitions in Appendix~\ref{sec:construction-details}).
This execution-based filtering reduces the candidate pool from 1,939 to 541 instances.




\subsection{Phase 5: Human Verification}
\label{sec:human-verification}
Following SWE-bench Verified~\citep{openai2024swebenchverified}, we perform a manual review to ensure benchmark quality. Two annotators independently examine each of the 541 candidate instances that passed execution-based filtering and resolve disagreements through discussion.
Review focuses on four aspects: \myNum{i}~Problem Statement Clarity to ensure that the GitHub issue provides sufficient context to understand the task; \myNum{ii}~Test Coverage Appropriateness by verifying that the tests correctly correspond to the issue being resolved; \myNum{iii}~Task Difficulty by labeling instances as easy, medium, or hard based on conceptual complexity, number of files modified, and required framework knowledge; and \myNum{iv}~Task Category by classifying each instance into one of the task types: Bug Fix, New Feature, or Feature Optimization.

Instances with severely underspecified issues or misaligned tests are excluded.
In total, 134 instances (24.8\%) fail manual verification, yielding a final benchmark of 407 instances. 
Difficulty and category labels are retained as metadata for analyzing model performance across task complexity levels and task types. More details are provided in Appendix~\ref{sec:annotation}.

\definecolor{rowgray}{gray}{0.92}
\begin{table*}[t]
\centering
\footnotesize
\setlength{\tabcolsep}{5pt}
\caption{Per-repository statistics for MobileDev-Bench. \#Files and \#LoC report codebase size at the base commit. \#~is the instance count. \#Tok is the average issue description length in tokens. Fix patch columns (\#L: changed lines, \#H: hunks, \#F: files) and test transition columns (F2P: Fail-to-Pass, N2P: None-to-Pass, P2P: Pass-to-Pass) report per-instance averages.}
\vspace{4pt}
\label{tab:instance-count}
\begin{tabular}{l|rr|rr|rrr|rrr}
\toprule
& \multicolumn{2}{c|}{\textbf{Codebase}} & \multicolumn{2}{c|}{\textbf{Instances}} & \multicolumn{3}{c|}{\textbf{Fix Patch (Avg.)}} & \multicolumn{3}{c}{\textbf{Tests (Avg.)}} \\
\cmidrule(lr){2-3} \cmidrule(lr){4-5} \cmidrule(lr){6-8} \cmidrule(lr){9-11}
\textbf{Repository} & \textbf{\#Files} & \textbf{\#LoC} & \textbf{\#} & \textbf{\#Tok} & \textbf{\#L} & \textbf{\#H} & \textbf{\#F} & \textbf{F2P} & \textbf{N2P} & \textbf{P2P} \\
\midrule
\rowcolor{rowgray}
\multicolumn{11}{l}{\textbf{Flutter (Dart)}} \\
\href{https://github.com/PalisadoesFoundation/talawa}{talawa} & 3,264 & 236k & 12 & 280 & 256.6 & 18.0 & 5.9 & 1.0 & 4.2 & 3.8 \\
\href{https://github.com/zulip/zulip-flutter}{zulip-flutter} & 514 & 141k & 51 & 325 & 254.9 & 18.4 & 7.9 & 1.2 & 53.7 & 49.9 \\
\rowcolor{rowgray}
\multicolumn{11}{l}{\textbf{Android Native (Java/Kotlin)}} \\
\href{https://github.com/AntennaPod/AntennaPod}{antennapod} & 1,143 & 115k & 9 & 1,567 & 348.1 & 35.7 & 12.9 & 0.0 & 15.7 & 0.0 \\
\href{https://github.com/commons-app/apps-android-commons}{apps-android-commons} & 1,439 & 150k & 10 & 548 & 1356.4 & 74.3 & 27.7 & 0.7 & 68.5 & 0.8 \\
\href{https://github.com/element-hq/element-x-android}{element-x-android} & 6,504 & 341k & 68 & 159 & 368.4 & 40.0 & 17.8 & 2.0 & 27.9 & 23.4 \\
\href{https://github.com/Futsch1/medTimer}{medtimer} & 556 & 41k & 8 & 152 & 454.0 & 33.2 & 17.2 & 0.5 & 4.6 & 0.6 \\
\href{https://github.com/JackEblan/Geto}{geto} & 451 & 23k & 1 & 156 & 172.0 & 33.0 & 5.0 & 0.0 & 52.0 & 0.0 \\
\href{https://github.com/LemmyNet/jerboa}{jerboa} & 337 & 122k & 4 & 218 & 12.8 & 1.8 & 1.2 & 4.0 & 11.8 & 1.0 \\
\href{https://github.com/mjaakko/NeoStumbler}{neostumbler} & 627 & 39k & 2 & 282 & 12.0 & 2.0 & 1.0 & 1.0 & 0.0 & 0.5 \\
\href{https://github.com/openhab/openhab-android}{openhab-android} & 1,173 & 52k & 8 & 228 & 584.1 & 13.5 & 8.1 & 0.6 & 10.9 & 5.6 \\
\href{https://github.com/PaulWoitaschek/Voice}{voice} & 649 & 37k & 4 & 259 & 273.0 & 10.0 & 7.5 & 0.3 & 4.8 & 0.5 \\
\href{https://github.com/streetcomplete/StreetComplete}{streetcomplete} & 5,534 & 629k & 18 & 411 & 625.4 & 52.4 & 27.1 & 0.6 & 68.3 & 7.1 \\
\href{https://github.com/thunderbird/thunderbird-android}{thunderbird-android} & 6,389 & 365k & 100 & 411 & 342.6 & 27.5 & 16.4 & 0.8 & 30.1 & 3.1 \\
\href{https://github.com/tuskyapp/Tusky}{tusky} & 1,593 & 169k & 12 & 318 & 46.5 & 4.8 & 2.4 & 1.4 & 18.5 & 7.8 \\
\href{https://github.com/wordpress-mobile/WordPress-Android}{wordpress-android} & 6,342 & 776k & 84 & 159 & 174.9 & 17.4 & 7.2 & 0.3 & 28.9 & 0.8 \\
\rowcolor{rowgray}
\multicolumn{11}{l}{\textbf{React-Native (TypeScript)}} \\
\href{https://github.com/artsy/eigen}{artsy/eigen} & 2,962 & 261k & 1 & 287 & 2311.0 & 116.0 & 18.0 & 3.0 & 0.0 & 0.0 \\
\href{https://github.com/Expensify/App}{expensify/app} & 7,923 & 2,652k & 2 & 547 & 429.0 & 42.0 & 17.5 & 1.5 & 1.0 & 2.0 \\
\href{https://github.com/NMF-earth/nmf-app}{NMF-earth/nmf-app} & 900 & 21k & 11 & 135 & 429.7 & 32.8 & 8.1 & 3.5 & 0.2 & 0.4 \\
\href{https://github.com/RocketChat/Rocket.Chat.ReactNative}{rocket.chat.reactnative} & 1,601 & 134k & 2 & 442 & 24.5 & 9.0 & 5.0 & 3.5 & 3.5 & 0.5 \\
\bottomrule
\end{tabular}
\vspace{-8pt}
\end{table*}




\section{Benchmark Characteristics}
\label{sec:characterization}

\subsection{Benchmark Statistics}
\label{subsec:statistics}

{\benchmark} comprises 407 manually verified instances drawn from 19 production mobile repositories spanning three ecosystems: Flutter (Dart), Android Native (Java/Kotlin), and React-Native (TypeScript).
Per-repository statistics are reported in Table~\ref{tab:instance-count}.
Repository sizes range from 337 to 7,923 source files (332K lines on average), reflecting the diversity of mature open-source codebases included in the benchmark.
Instance counts vary from 1 to 101 per repository, with the four largest contributors being thunderbird-android~(100), wordpress-android~(84), element-x-android~(68), and zulip-flutter~(51), together accounting for 74.4\% of the dataset.

Table~\ref{tab:dataset-summary-stats} reports aggregate statistics across the dataset. Issue descriptions average 308 tokens (median 173), with a long tail reaching 11,501 tokens for the most detailed reports, reflecting the context-rich nature of app bug reports that typically include reproduction steps, platform versions, and device-specific conditions.
None-to-Pass is the dominant validation signal: 348 of 407 instances (85.5\%) yield at least one N2P transition, with a mean of 31.1 and a median of 11. N2P transitions arise from two sources: tests that did not exist at the base commit and are introduced by the fix, and tests that were present but could not be executed due to compilation failures at the base commit, a common occurrence in statically typed languages such as Java and Kotlin where the fix itself is required to restore compilability.
Additionally, 147 instances (36.1\%) yield at least one Fail-to-Pass transition (mean 1.1), indicating cases where pre-existing tests directly captured the regression.
The high N2P mean is driven by repositories with large test suites (e.g., zulip-flutter avg.\ 53.7 N2P per instance), where the fix enables previously non-executable test modules or introduces newly added tests.

\begin{table}[t]
\footnotesize
\setlength{\tabcolsep}{4pt}
\centering
\caption{Summary statistics for {\benchmark}.}
\begin{tabular}{l rr rr rrr rrr}
\toprule
 & \multicolumn{1}{c}{\textbf{Issue}} 
 & \multicolumn{2}{c}{\textbf{Codebase}} 
 & \multicolumn{3}{c}{\textbf{Fix Patch}} 
 & \multicolumn{3}{c}{\textbf{Tests}} \\
\cmidrule(lr){2-2}\cmidrule(lr){3-4}\cmidrule(lr){5-7}\cmidrule(lr){8-10}
 & Tokens & \#Files & \#Lines\,(K) & \#Lines & \#Files & \#Hunks & F2P & N2P & P2P \\
\midrule
Mean   &    308 & 4{,}605 & 389 &  334.6 & 12.9 & 27.5 &  1.1 & 31.1 & 11.9 \\
Median &    173 & 6{,}342 & 365 &   93 &  5   & 12   &  0   & 11   &  0   \\
Max    & 11{,}501 & 7{,}923 & 2{,}652 & 11{,}915 & 237 & 592 & 30 & 375 & 267 \\
\bottomrule
\end{tabular}
\label{tab:dataset-summary-stats}
\vspace{-10pt}
\end{table}

\begin{figure}[htb]
  \centering
  \includegraphics[width=0.8\linewidth]
  {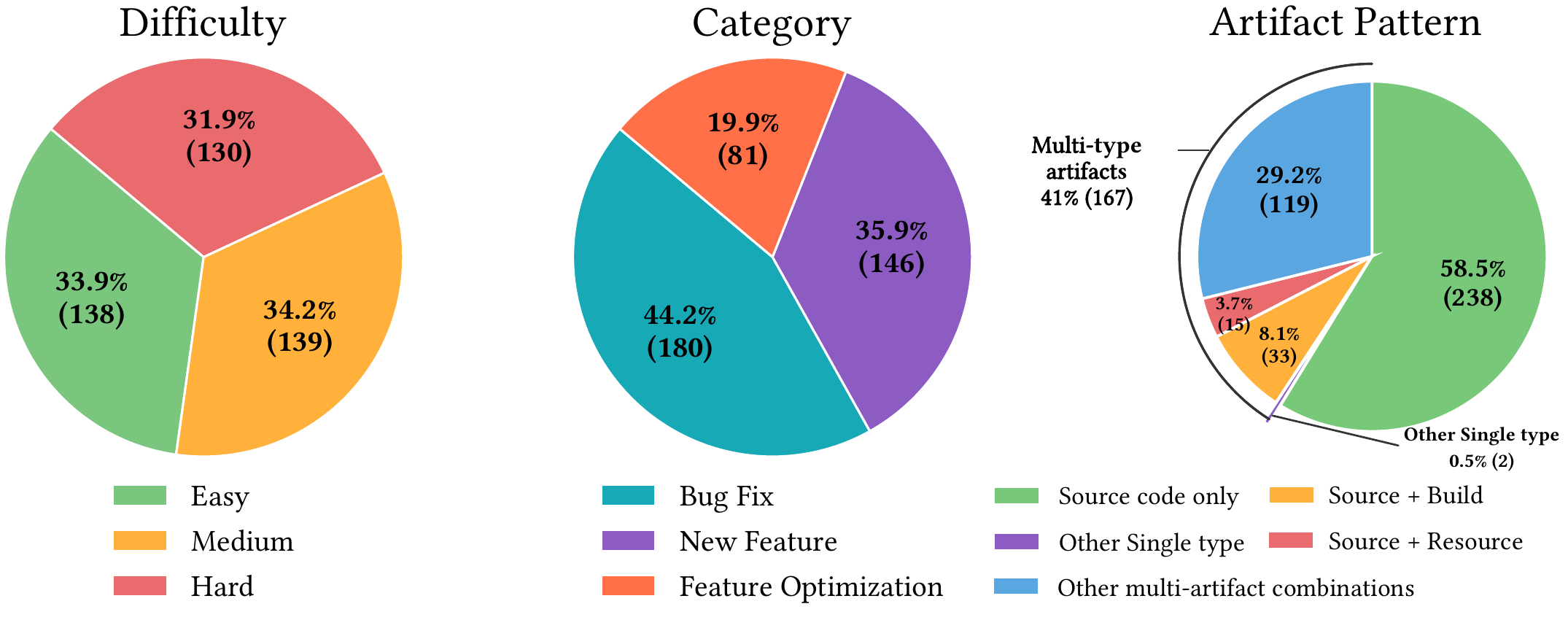}
  \vspace{-2pt}
  \caption{
  Distribution of MobileDev-Bench tasks by difficulty, category, and artifact pattern. Tasks are balanced across difficulty tiers, with bug fixes as the largest category. A significant number of tasks (41.0\%) require coordinated changes across multiple artifact types. 
  }
  \label{fig:difficulty-distribution}
  \vspace{-5pt}
\end{figure}

Figure~\ref{fig:difficulty-distribution} shows the distribution of instances in {\benchmark} across difficulty and task category. 
Difficulty is broadly balanced across easy (under 15 min, 33.9\%), medium (15 min to 1 hr, 34.2\%), and hard (over 1 hr, 31.9\%), ensuring the benchmark exercises a range of problem complexities.
By task category, 44.2\% of instances are bug fixes, 35.9\% are new features, and 19.9\% are feature optimizations.

\begin{figure*}[t]
    \centering
    \footnotesize
    \includegraphics[width=0.9\linewidth]{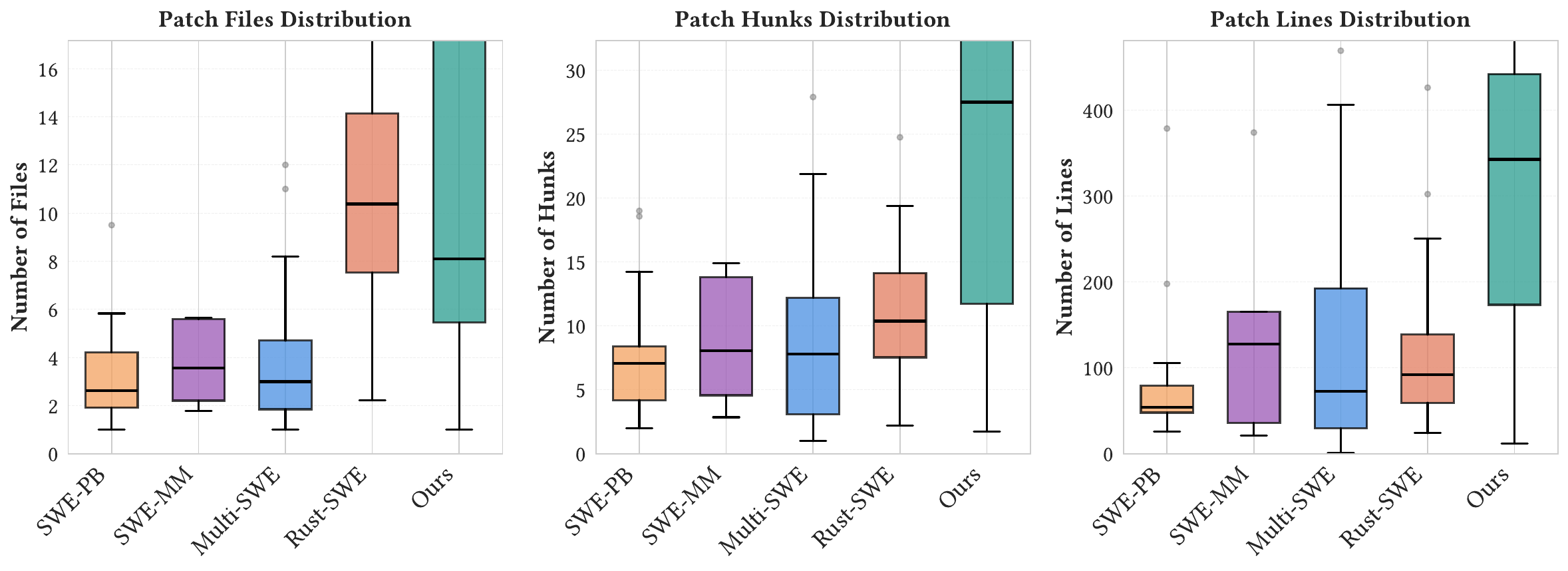}
    \vspace{-10pt}
    \caption{Distribution of patch complexity metrics (files modified, hunks, changed lines) across five benchmarks (SWE-PB: SWE-PolyBench; SWE-MM: SWE-bench Multimodal; Multi-SWE: Multi-SWE-bench; Rust-SWE: Rust-SWE-bench; Ours: {\benchmark}). Each data point is the per-repository average. Y-axes are capped at the 90th percentile for visual clarity.}
    \label{fig:patch-complexity}
    \vspace{-10pt}
\end{figure*}

Figure~\ref{fig:patch-complexity} shows that fix patches in {\benchmark} substantially exceed those in comparable benchmarks. Compared with SWE-PolyBench (closest in repository count), fixes are 3.1$\times$ larger in files, 3.9$\times$ in hunks, and 6.3$\times$ in changed lines, reflecting the coordinated nature of mobile fixes that span source, resources, and build configuration (see per-benchmark breakdown in Appendix~\ref{sec:benchmarks-comparison}).

\subsection{Benchmark Features}
\label{subsec:benchmark-features}
{\benchmark} captures characteristics of mobile app development that are absent from existing repository-level benchmarks. In particular, several fixes span multiple artifact types and may involve more than one language within a single patch.


\noindent\textbf{Multi-Artifact Fix Patterns.}
Mobile app fixes often require coordinated changes across heterogeneous artifact types. We classify modified files in each patch into eight categories:
\begin{itemize}[itemsep=1pt,topsep=2pt,leftmargin=25pt]
  \item \textbf{Source:} Application source code
        (\texttt{.kt}, \texttt{.java}, \texttt{.dart}, \texttt{.ts}, \texttt{.tsx},
        \texttt{.js}).
  \item \textbf{Resource:} UI and media assets
        (\texttt{res/} subtree, images, drawables, layouts).
  \item \textbf{i18n:} Localized string resources
        (Android \texttt{values-*/strings.xml}, Flutter \texttt{.arb}).
  \item \textbf{Build:} Build scripts         (\texttt{build.gradle}, \texttt{libs.versions.toml}, \texttt{*.kts}).
  \item \textbf{Manifest:} App-level declaration files
        (\texttt{AndroidManifest.xml}, \texttt{pubspec.yaml}).
  \item \textbf{Config:} CI/CD workflows and configuration
        (GitHub Actions, linter and formatter configs).
  \item \textbf{Docs:} Project documentation
        (Markdown, changelog fragments, license files).
  \item \textbf{Other:} Files not matched by any of the above.
\end{itemize}


Source files appear in nearly all instances (99.5\%). However, many fixes also require non-source changes: 20.4\% modify resources, 17.2\% modify localization files, 18.4\% modify build files, and 3.7\% modify manifest files. As summarized in Figure~\ref{fig:difficulty-distribution}, 41.0\% of instances modify two or more artifact types and 20.9\% modify three or more, demonstrating that cross-artifact coordination is a common property of mobile app fixes.
Full combination-level counts with per-language breakdowns are reported in Table~\ref{tab:artifact-combinations} in the appendix.


\noindent\textbf{Intra-Patch Language Diversity.}
Fix patches also frequently span multiple languages. 
We identify cross-language instances by examining the set of source-language file extensions modified within each fix patch. In total, 20.7\% of instances modify source files written in more than one language.
The most common pattern is Kotlin--Java co-modification, representing 19.8\% of all instances and 25.2\% of Kotlin-based fixes. This reflects the ongoing migration of Android projects from Java to Kotlin, with co-modification rates varying widely across repositories depending on their stage of migration: projects retaining residual Java code exhibit rates above 40\%, while repositories that have fully transitioned to Kotlin show none. A second cross-language pattern occurs in React-Native repositories, where TypeScript--JavaScript co-modification appears in 18.8\% of React-Native instances, reflecting gradual TypeScript adoption in existing JavaScript codebases.

\section{Experimental Setup}
\label{sec:evaluation}

We adopt Agentless~\citep{xia2024agentless} as our primary evaluation framework and extend it to support Java, Kotlin, TypeScript, and Dart by replacing its Python-specific AST parser with tree-sitter~\citep{tree-sitter}. 
Additional details are provided in Appendix~\ref{app:lang-adaptation}.

\noindent\textbf{Retrieval Settings.}
We evaluate model performance under two retrieval settings.
\myNum{i}~Automated retrieval: the model must identify which files to modify before generating a patch given the issue description. 
This end-to-end setting reflects the realistic repair scenario.
\myNum{ii}~Oracle retrieval~\citep{jimenez2024swebench}: the ground-truth set of files to modify is provided to the model in addition to the issue description. 
While less realistic, it can provide an upper bound on repair performance achievable under ideal file-level localization conditions.
In both settings, generated patches are applied to the base commit and validated by executing the repository test suite within our Docker-based~\citep{merkel2014docker} containerized environment.

\noindent\textbf{Metrics.}
Following prior work~\citep{jimenez2024swebench, xia2024agentless}, we report two metrics evaluated under both retrieval settings.
\myNum{i}~\emph{Resolution Rate}: the percentage of tasks where the generated patch passes the full test suite.
\myNum{ii}~\emph{Localization Metrics}: file-level Precision, Recall, and F1 (macro-averaged across tasks), measuring how accurately the model's patch covers the ground-truth modified files.
Metric definitions are provided in Appendix~\ref{app:appendix-metrics}.


\noindent\textbf{Models.}
We evaluate four frontier code-capable models spanning proprietary and open-weight: Claude Sonnet 4.5~\citep{anthropic2025claude}, GPT-5.2~\citep{openai2025gpt52}, Gemini 2.5 Flash~\citep{comanici2025gemini25pushingfrontier}, and Qwen-3-Coder~\citep{qwen2025qwen3coder}.
To ensure deterministic patch generation and fair comparison, we follow the original Agentless evaluation protocol.

\section{Results}
\label{sec:results}
\subsection{Resolution Performance}
Table~\ref{tab:resolution-rate-difficulty} reports resolution rates under both retrieval settings.
Under automated retrieval, overall performance remains low, ranging from 3.23\% to 4.23\% across models. 
Performance degrades with task difficulty: easy tasks achieve 8.0\%--8.5\%, medium tasks drop to 1.5\%--4.4\%, and hard tasks are not resolved by any model.

\begin{figure*}[hbt]
\centering
\vspace{-5pt}
\includegraphics[width=0.8\textwidth]{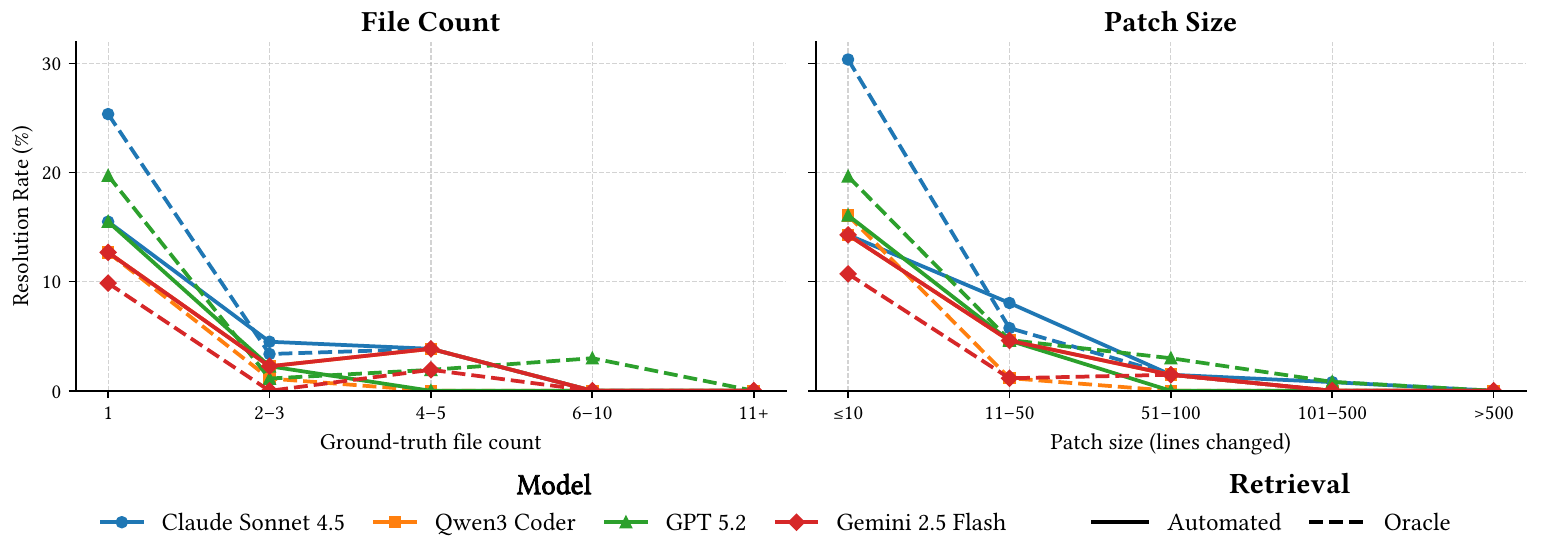}
\vspace{-5pt}
\caption{Resolution rates by ground-truth file count and patch size. Single-file tasks achieve 13--16\% (Automated) and 10--25\% (Oracle); rates collapse for multi-file and large-patch tasks.}
\vspace{-5pt}
\label{fig:resolution-combined}
\end{figure*}

Artifact diversity further exacerbates this trend. 
For single-type artifacts, resolution rates remain modest (5.0\%--6.7\%), but performance collapses to 0\%--0.6\% when multiple artifact types are involved. 
As shown in Figure~\ref{fig:resolution-combined}, resolution rates also decline steeply with structural complexity: single-file tasks achieve 12.7\%–15.5\%, whereas performance approaches zero for tasks involving 6+ files or patches exceeding 50 lines.

Under oracle retrieval, Claude Sonnet 4.5 and GPT-5.2 improve to 5.69\% and 4.53\% overall, respectively. 
In contrast, Qwen3 Coder and Gemini 2.5 Flash degrade relative to automated retrieval (from 3.33\% to 2.57\% and 3.24\% to 1.98\%, respectively). 
Our manual analysis of the generated patches shows over-editing: when given a full set of files to modify, models often introduce unnecessary changes beyond the required fix, leading to regressions. 
Claude Sonnet 4.5 and GPT-5.2 exhibit this behavior to a lesser extent, explaining their positive gains under oracle retrieval. Appendix~\ref{app:appendix_oracle_regression} provides analysis and examples of over-editing.

\begin{table}[t!]
\centering
\footnotesize
\setlength{\tabcolsep}{3pt}
\caption{Resolution rates (\%) by difficulty level and artifact diversity under automated and oracle retrieval. Results show that resolution rates decline as difficulty and artifact complexity increase.}
\label{tab:resolution-rate-difficulty}
\begin{tabular}{llcccccccc}
\toprule
& & \multicolumn{4}{c}{\textbf{Automated Retrieval}} & \multicolumn{4}{c}{\textbf{Oracle Retrieval}} \\
\cmidrule(lr){3-6}\cmidrule(lr){7-10}
\textbf{Difficulty / artifact types} & \textbf{Tasks} & \textbf{Claude} & \textbf{Qwen} & \textbf{GPT} & \textbf{Gemini} & \textbf{Claude} & \textbf{Qwen} & \textbf{GPT} & \textbf{Gemini} \\
\midrule
Easy     &  138 (33.9\%)  &  8.0 &  8.5 &  8.0 &  8.0 & 13.0 &  6.7 & 10.9 &  4.3 \\
Medium    &  139 (34.2\%) &  4.4 &  1.5 &  1.5 &  1.5 &  2.9 &  0.7 &  2.2 &  1.4 \\
Hard      &  130 (31.9\%)  &  0.0 &  0.0 &  0.0 &  0.0 &  0.8 &  0.0 &  0.0 &  0.0 \\
\midrule
Single-type artifact & 240 (59\%) & 6.7 & 5.0 & 5.4 & 5.0 &  9.6 & 4.2 & 7.1 & 3.3 \\
Multi-type artifact & 167 (41\%) & 0.6 & 0.6 & 0.0 & 0.6 &  0.0 & 0.0 & 0.6 & 0.0 \\
\midrule
Overall   &  407 & 4.23 & 3.33 & 3.23 & 3.24 & 5.69 & 2.57 & 4.53 & 1.98 \\
\bottomrule
\end{tabular}
\vspace{-5pt}
\end{table}

\subsection{Resolution Characteristics}


\noindent\textbf{Problem Difficulty.}
Resolution rates decline with task difficulty under both retrieval settings. 
Specifically, hard tasks are not resolved by any model under automated retrieval, and only Claude Sonnet 4.5 achieves a non-zero rate of 0.8\% under oracle, indicating that difficulty stratification in {\benchmark} captures a wide and challenging range of solution complexity.

\noindent\textbf{Artifact Type Diversity.}
Resolution success is strongly associated with artifact diversity. Single-artifact tasks constitute 59\% of the benchmark and achieve resolution rates of 5.0\%--6.7\% under automated retrieval, with Claude Sonnet 4.5 and GPT-5.2 rising to 9.6\% and 7.1\% under oracle retrieval. 
In contrast, multi-artifact tasks constitute 41\% of the benchmark but reach at most 0.6\% under either retrieval setting across all models, indicating that current models struggle to coordinate fixes across heterogeneous artifact types.


\noindent\textbf{File Count Effects.}
As shown in Figure~\ref{fig:resolution-combined} (left), resolution rates drop steeply as the number of files to modify increases. 
Single-file tasks achieve 12.7\%–15.5\% under automated retrieval, with oracle rates reaching up to 25.4\% for some models. 
Rates collapse to near zero for tasks requiring 4 or more files, with 6–10 file tasks yielding 0\% under automated retrieval across all models and remaining near zero even under oracle retrieval, indicating that multi-file coordination is a major source of difficulty in mobile app issue resolution.


\noindent\textbf{Patch Size Effects.}
As shown in Figure~\ref{fig:resolution-combined} (right), resolution rates decline with patch size across both retrieval settings. Tasks requiring at most 10 lines achieve 14.3\%–16.1\% under automated retrieval, while rates collapse to near zero beyond 50 lines and larger patches. 
Both settings converge toward zero for large patches, indicating that patch generation itself fails as fix complexity grows, independent of localization quality.

\begin{figure}[t!]
\centering
\includegraphics[width=0.95\textwidth]{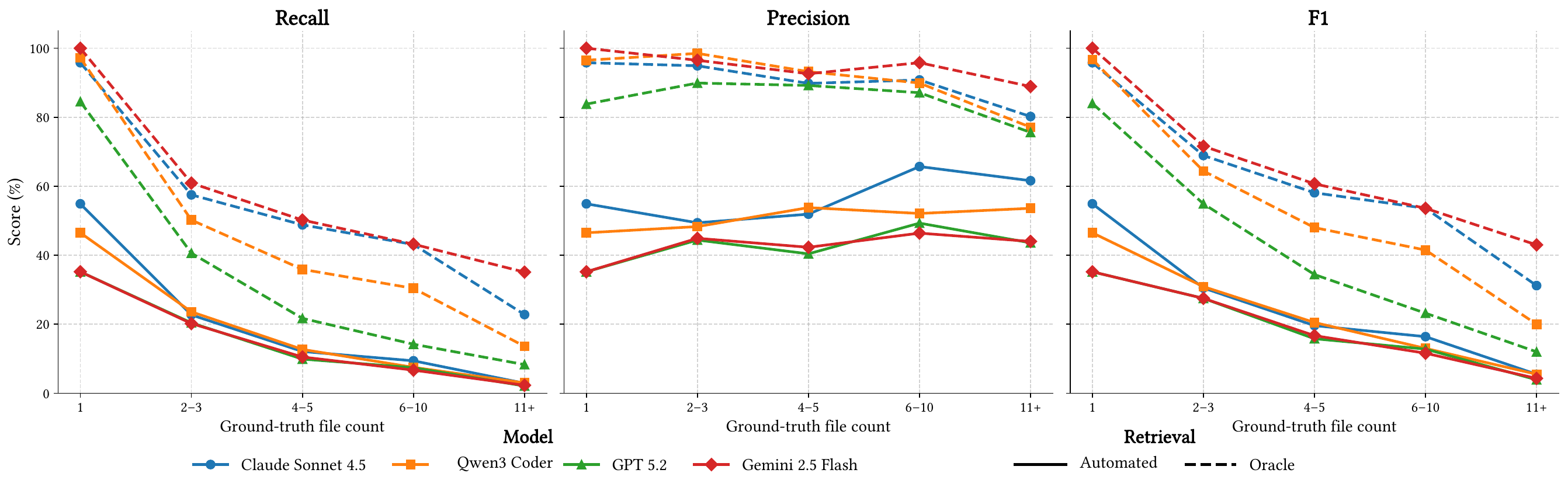}
\vspace{-8pt}
\caption{Recall and F1 scores decrease across all evaluated models as the number of modified files increases. Oracle retrieval consistently outperforms Automated retrieval across complexity levels.}
\label{fig:recall-filecount}
\vspace{-8pt}
\end{figure}

\subsection{Fault Localization Performance}
Figure~\ref{fig:recall-filecount} reports Recall, Precision, and F1 under both retrieval settings as a function of ground-truth file count. Under automated retrieval, overall recall ranges from 13.8\% to 18.6\%, with precision substantially higher (42.9\%--57.2\%), indicating that when models do retrieve relevant context they tend to modify appropriate files, but they miss a large fraction of required modification files. Recall declines with file count, from 35--55\% on single-file tasks to 2--3\% on tasks with 11 or more files.

Under oracle retrieval, recall rises substantially, reaching 84.5--100\% on single-file tasks, confirming that models can modify the correct files when those files are provided directly. However, oracle resolution rates peak at only 5.69\% (Table~\ref{tab:resolution-rate-difficulty}), indicating that models edit the correct files but still fail to produce patches that pass the test suite. 
Recall under oracle also degrades with file count, falling to 8.3--35.1\% on tasks with 11 or more files, pointing to an additional difficulty in generating changes across all provided files simultaneously. 
Comprehensive resolution rate breakdowns, per-model analysis, and additional resolution characteristics are provided in Appendix~\ref{sec:appendix_evaluation_results}.

\section{Conclusion}
\label{sec:conclusion}

We introduced {\benchmark}, a benchmark of 407 verified issue-resolution tasks from 19 production Android Native, React Native, and Flutter repositories. 
Evaluation of four frontier LLMs yields resolution rates of 3.23\%--4.23\% under automated retrieval and at most 5.69\% under oracle retrieval, well below rates reported on existing benchmarks. 
File-level recall of 13.8\%--18.6\% under automated retrieval declines steeply with task complexity, limiting resolution rates. 
Under oracle retrieval, recall rises to 31.4\%--55.4\% overall and 84.5\%--100\% on single-file tasks, yet resolution rates remain below 6\%, indicating that correctly identifying files to modify is necessary but not sufficient for successful repair. 
These results suggest that progress on mobile app issue resolution requires advances in both multi-file dependency reasoning and patch generation under heterogeneous artifact constraints.
{\benchmark} provides a reproducible foundation for future research on AI-assisted mobile app development.

\bibliographystyle{abbrvnat}
\bibliography{references}

\newpage
\appendix

\section{Additional Construction Details}
\label{sec:construction-details}

\begin{table*}[t]
\centering
\small
\setlength{\tabcolsep}{5pt}
\caption{Summary and licenses for all GitHub repositories represented in {\benchmark}.}
\label{tab:summary-licenses}
\begin{tabular}{p{4.2cm}p{6cm}l}
\toprule
\textbf{Repository} & \textbf{Summary} & \textbf{License} \\
\midrule
palisadoesfoundation/talawa & Community organization management & GNU GPLv3 \\
zulip/zulip-flutter & Zulip mobile apps for Android and iOS & Apache 2.0 \\
antennapod/antennapod & Podcast manager for Android & GNU GPLv3 \\
commons-app/apps-android-commons & Wikimedia Commons Android uploader & Apache 2.0 \\
element-hq/element-x-android & Android Matrix messenger (Jetpack Compose) & GNU AGPLv3 \\
futsch1/medtimer & Medication reminder app & MIT \\
jackeblan/geto & Per-app device settings manager & GNU GPLv3 \\
lemmynet/jerboa & Native Android client for Lemmy & GNU AGPLv3 \\
mjaakko/neostumbler & OpenStreetMap geolocation stumbler & MIT \\
openhab/openhab-android & openHAB client for Android & EPL 2.0 \\
paulwoitaschek/voice & Minimalist audiobook player & GNU GPLv3 \\
streetcomplete/streetcomplete & Easy-to-use OpenStreetMap editor & GNU GPLv3 \\
thunderbird/thunderbird-android & Thunderbird for Android (fka K-9 Mail) & Apache 2.0 \\
tuskyapp/tusky & Android client for Mastodon & GNU GPLv3 \\
wordpress-mobile/wordpress-android & WordPress for Android & GNU GPLv2 \\
artsy/eigen & Artsy's iOS/Android art marketplace & MIT \\
expensify/app & Financial collaboration platform & MIT \\
NMF-earth/nmf-app & Carbon footprint tracker for daily life & GNU GPLv3 \\
rocketchat/rocket.chat.reactnative & Secure communications platform & MIT \\
\bottomrule
\end{tabular}
\end{table*}

This appendix provides supplementary details for the construction pipeline described in Section~\ref{sec:benchmark}.

\subsection{Repository Selection Filters}
\label{sec:app-selection-filters}

The following five filters are applied during Phase~1 (App Selection) to determine benchmark viability:
\begin{enumerate}
  \renewcommand{\theenumi}{\roman{enumi}}
  \renewcommand{\labelenumi}{(\theenumi)}
  \item A minimum of 400 GitHub stars to ensure community adoption and sustained development.
  \item Recent maintenance within the past six months, verified through commit activity, merged pull requests, and responsive issue management.
  \item A permissive open-source license (e.g., Apache 2.0, MIT, or GPL variants) allowing research use.
  \item An issue tracker hosted on GitHub, enabling reliable retrieval of issue--PR mappings and contextual information.
  \item The latest commit must build and execute its test suite in a clean environment. For Android Native apps, we install the required JDK version (11/17/21) and run Gradle build and test tasks. For React Native apps, we verify Node.js compatibility and run tests via npm or yarn. For Flutter apps, we confirm Flutter SDK compatibility and execute the test framework. Repositories that fail this check are excluded.
\end{enumerate}

\subsection{Test File Identification Patterns}
\label{sec:test-patterns}

During Phase~2 (PR Collection and Filtering), test files are identified using the following framework-specific naming conventions:
\begin{itemize}
  \item \textbf{Android Native:} \texttt{*Test.kt}, \texttt{*Test.java}, or files under \texttt{androidTest/} and \texttt{test/} source sets.
  \item \textbf{React Native:} \texttt{*.test.ts}, \texttt{*.test.tsx}, \texttt{*.test.js}, or \texttt{*.spec.ts} files.
  \item \textbf{Flutter:} \texttt{*\_test.dart} files or any file under the \texttt{test/} directory.
\end{itemize}

\subsection{Repository Composition}

{\benchmark} draws from 19 open-source Android and cross-platform mobile repositories spanning utility apps, media players, productivity tools, and developer tooling.
Table~\ref{tab:summary-licenses} lists all 19 repositories with a brief description of their purpose and their open-source license.
The selection prioritises projects with active CI pipelines, high test coverage, and sustained PR activity to ensure a sufficient supply of execution-validated instances.

\subsection{Pipeline Conversion Statistics}

Table~\ref{tab:conversion-stats} reports per-repository counts at each stage of the conversion pipeline: PRs crawled, attribute-filtered candidates, execution-validated instances, and manually verified benchmark instances.
Overall, the pipeline reduces 74,195 crawled PRs to 407 verified instances (0.5\% retention rate).
The primary bottleneck is attribute filtering (Phase~1), which discards PRs that lack an explicit issue link, do not modify test files, or touch more files than the complexity threshold permits.
Execution validation (Phase~4) constitutes the second largest reduction, as a substantial fraction of candidates fail to build or produce non-deterministic test outcomes.

\begin{table*}[htbp]
\centering
\small
\setlength{\tabcolsep}{5pt}
\caption{Benchmark repository instance conversion pipeline showing progression from collected pull requests through attribute-based filtering, execution-based validation, and manual verification.}
\label{tab:conversion-stats}
\begin{tabular}{l | r r r r}
\toprule
 & \makecell{\textbf{PRs}\\\textbf{Crawled}} & \textbf{Filtered} & \textbf{Validated} & \textbf{Verified} \\
\midrule
palisadoesfoundation/talawa & 1,743 & 212\textsubscript{\textcolor{red}{$\downarrow$1531}} & 13\textsubscript{\textcolor{red}{$\downarrow$199}} & 12\textsubscript{\textcolor{red}{$\downarrow$1}} \\
zulip/zulip-flutter & 1,051 & 96\textsubscript{\textcolor{red}{$\downarrow$955}} & 51\textsubscript{\textcolor{red}{$\downarrow$45}} & 51\textsubscript{\textcolor{gray}{.0}} \\
antennapod/antennapod & 2,031 & 89\textsubscript{\textcolor{red}{$\downarrow$1942}} & 11\textsubscript{\textcolor{red}{$\downarrow$78}} & 9\textsubscript{\textcolor{red}{$\downarrow$2}} \\
commons-app/apps-android-commons & 1,477 & 200\textsubscript{\textcolor{red}{$\downarrow$1277}} & 17\textsubscript{\textcolor{red}{$\downarrow$183}} & 10\textsubscript{\textcolor{red}{$\downarrow$7}} \\
element-hq/element-x-android & 3,899 & 303\textsubscript{\textcolor{red}{$\downarrow$3596}} & 77\textsubscript{\textcolor{red}{$\downarrow$226}} & 68\textsubscript{\textcolor{red}{$\downarrow$9}} \\
futsch1/medtimer & 643 & 25\textsubscript{\textcolor{red}{$\downarrow$618}} & 12\textsubscript{\textcolor{red}{$\downarrow$13}} & 8\textsubscript{\textcolor{red}{$\downarrow$4}} \\
jackeblan/geto & 195 & 20\textsubscript{\textcolor{red}{$\downarrow$175}} & 6\textsubscript{\textcolor{red}{$\downarrow$14}} & 1\textsubscript{\textcolor{red}{$\downarrow$5}} \\
lemmynet/jerboa & 1,048 & 15\textsubscript{\textcolor{red}{$\downarrow$1033}} & 5\textsubscript{\textcolor{red}{$\downarrow$10}} & 4\textsubscript{\textcolor{red}{$\downarrow$1}} \\
mjaakko/neostumbler & 681 & 20\textsubscript{\textcolor{red}{$\downarrow$661}} & 5\textsubscript{\textcolor{red}{$\downarrow$15}} & 2\textsubscript{\textcolor{red}{$\downarrow$3}} \\
openhab/openhab-android & 1,607 & 27\textsubscript{\textcolor{red}{$\downarrow$1580}} & 12\textsubscript{\textcolor{red}{$\downarrow$15}} & 8\textsubscript{\textcolor{red}{$\downarrow$4}} \\
paulwoitaschek/voice & 1,364 & 10\textsubscript{\textcolor{red}{$\downarrow$1354}} & 6\textsubscript{\textcolor{red}{$\downarrow$4}} & 4\textsubscript{\textcolor{red}{$\downarrow$2}} \\
streetcomplete/streetcomplete & 1,167 & 96\textsubscript{\textcolor{red}{$\downarrow$1071}} & 27\textsubscript{\textcolor{red}{$\downarrow$69}} & 18\textsubscript{\textcolor{red}{$\downarrow$9}} \\
thunderbird/thunderbird-android & 2,728 & 245\textsubscript{\textcolor{red}{$\downarrow$2483}} & 110\textsubscript{\textcolor{red}{$\downarrow$135}} & 100\textsubscript{\textcolor{red}{$\downarrow$10}} \\
tuskyapp/tusky & 2,110 & 28\textsubscript{\textcolor{red}{$\downarrow$2082}} & 19\textsubscript{\textcolor{red}{$\downarrow$9}} & 12\textsubscript{\textcolor{red}{$\downarrow$7}} \\
wordpress-mobile/wordpress-android & 4,199 & 438\textsubscript{\textcolor{red}{$\downarrow$3761}} & 140\textsubscript{\textcolor{red}{$\downarrow$298}} & 84\textsubscript{\textcolor{red}{$\downarrow$56}} \\
artsy/eigen & 10,020 & 3\textsubscript{\textcolor{red}{$\downarrow$10017}} & 2\textsubscript{\textcolor{red}{$\downarrow$1}} & 1\textsubscript{\textcolor{red}{$\downarrow$1}} \\
expensify/app & 34,301 & 30\textsubscript{\textcolor{red}{$\downarrow$34271}} & 8\textsubscript{\textcolor{red}{$\downarrow$22}} & 2\textsubscript{\textcolor{red}{$\downarrow$6}} \\
NMF-earth/nmf-app & 181 & 23\textsubscript{\textcolor{red}{$\downarrow$158}} & 15\textsubscript{\textcolor{red}{$\downarrow$8}} & 11\textsubscript{\textcolor{red}{$\downarrow$4}} \\
rocketchat/rocket.chat.reactnative & 3,750 & 59\textsubscript{\textcolor{red}{$\downarrow$3691}} & 5\textsubscript{\textcolor{red}{$\downarrow$54}} & 2\textsubscript{\textcolor{red}{$\downarrow$3}} \\
\midrule
\textbf{Total} & 74,195 & 1,939\textsubscript{\textcolor{red}{$\downarrow$72256}} & 541\textsubscript{\textcolor{red}{$\downarrow$1398}} & 407\textsubscript{\textcolor{red}{$\downarrow$134}} \\
\bottomrule
\end{tabular}
\end{table*}

\subsection{Environment Dockerfile Generation and Remediation}

After generating a candidate Dockerfile from parsed configuration (as described in Phase~3, Section~\ref{sec:benchmark}), we run a full build-and-test cycle to validate correctness. Failed builds are manually inspected; common fixes include adding missing system packages (e.g., \texttt{libssl-dev} for React Native native modules), pinning a breaking transitive dependency, or correcting a misidentified JDK version. Repositories requiring more than three remediation rounds are excluded.

For Android projects, GitHub Actions workflow files (\texttt{.github/workflows/*.yml}) often provide the most precise environment specification, pinning exact Gradle wrapper versions, JDK versions, and Android SDK command-line tool versions that are absent from \texttt{build.gradle} or \texttt{gradle.properties}.

\subsection{\texttt{NONE} State and None-to-Pass Semantics}

The \texttt{NONE} test state arises when a test class references types or methods introduced only by the fix patch; the class is absent from the compiled base commit, so the test runner cannot locate or execute it.
This is distinct from \texttt{FAIL}, where the test compiles and runs but produces an incorrect result.
A \texttt{NONE}$\to$\texttt{PASS} (N2P) transition therefore indicates a newly introduced test that becomes executable only once the fix is applied.
Repositories that follow a test-driven development workflow may exhibit exclusively N2P transitions with zero F2P counts, because their test patches introduce test classes that are entirely absent from the base commit.

\subsection{Multi-Module Gradle Test Targeting}

For multi-module Android repositories, we execute module-specific Gradle tasks (e.g., \texttt{:feature:login:testDebugUnitTest}) rather than the top-level \texttt{test} task, targeting only modules containing files touched by each PR's test patch.
Relevant modules are identified by mapping modified test files to their enclosing Gradle module via \texttt{settings.gradle} declarations.
We default to the \texttt{Debug} build variant, which is the standard configuration used by the CI pipelines of the Android repositories in the benchmark.

\section{Manual Validation Details}
\label{sec:annotation}

\begin{figure}[]
\centering
\includegraphics[width=\columnwidth]{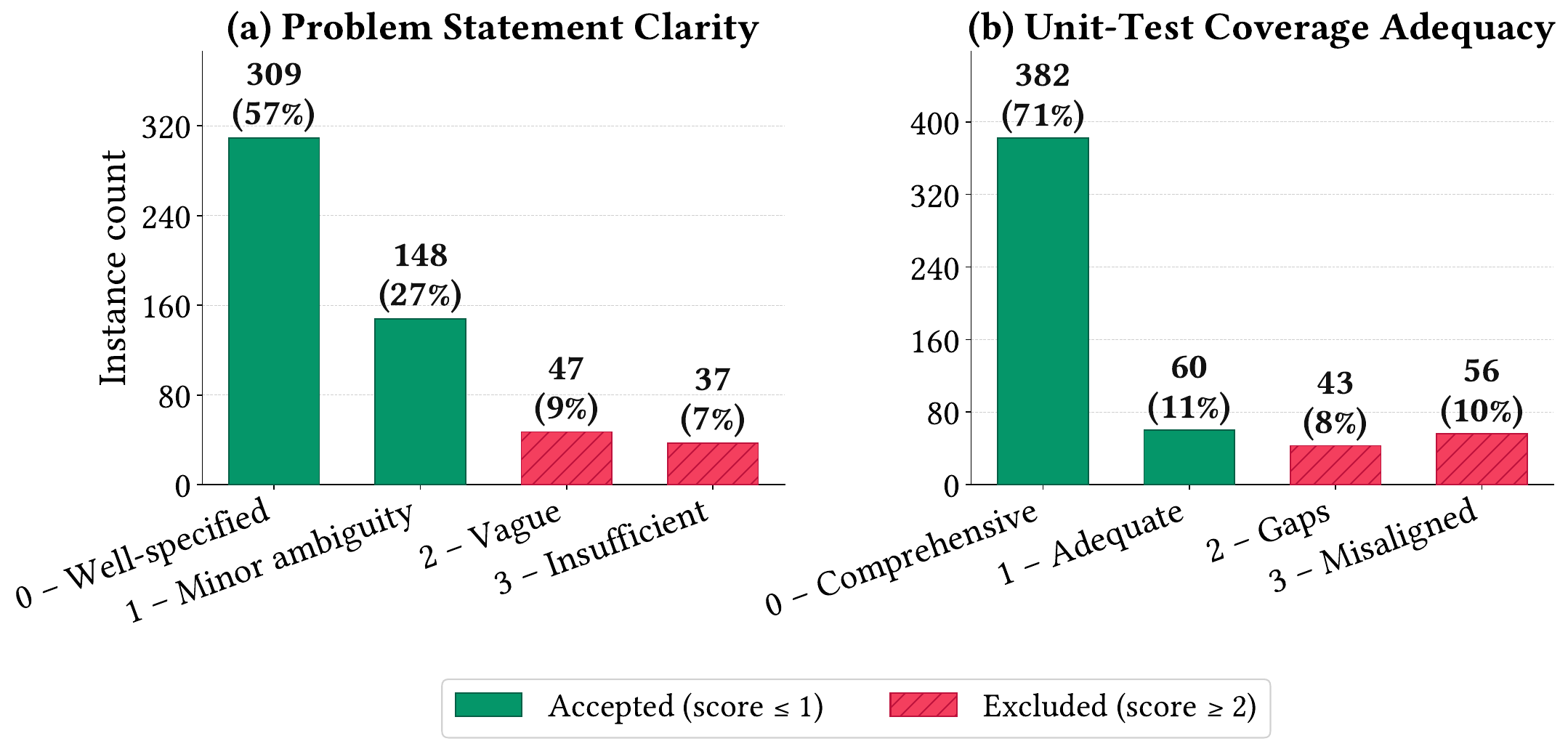}
\caption{Distribution of manual annotation scores for
  \textbf{(a)} problem-statement clarity and
  \textbf{(b)} unit-test coverage adequacy,
  across all 541 annotated candidate instances.
  Green bars (scores 0--1) indicate instances accepted into the benchmark;
  hatched red bars (scores 2--3) indicate excluded instances.
  The majority of exclusions arise from vague issue descriptions (score 2/3
  on dimension~a) or tests that impose undue implementation constraints
  (score 2/3 on dimension~b).}
\label{fig:annotation-quality}
\end{figure}

\begin{figure}[]
\centering
\includegraphics[width=\columnwidth]{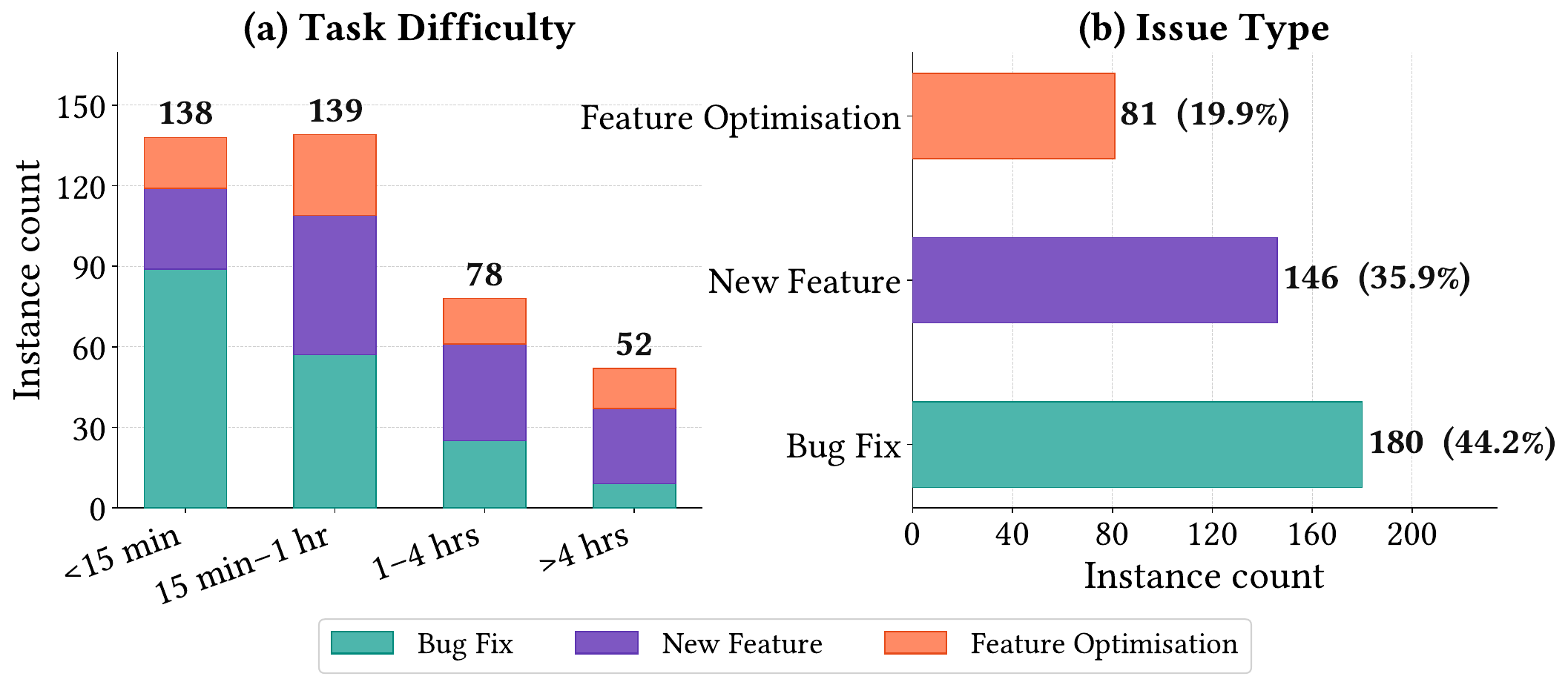}
\caption{Task characteristics of the 407 verified benchmark instances.
  \textbf{(a)} Difficulty distribution stacked by issue type, using the four
  time-based difficulty labels assigned during annotation.
  \textbf{(b)} Overall issue-type breakdown.
  The benchmark spans the full difficulty range and covers all three major
  categories of mobile development activity.}
\label{fig:task-characteristics}
\end{figure}

Following SWE-bench-Verified~\citep{openai2024swebenchverified}, we perform human review of each instance that passes execution-based filtering.
The review criteria are described in Section~\ref{sec:benchmark} (Phase~5).
Figures~\ref{fig:annotation-quality} and~\ref{fig:task-characteristics} visualise the score distributions and dataset composition, respectively.

\subsection{Annotation Results}

Among the 541 annotated candidates, 309 instances (57.1\%) have unambiguously well-specified problem statements (score~0) and 148 (27.4\%) have minor ambiguity but a clear interpretation (score~1).
The remaining 84 (15.5\%) are excluded by problem-statement score: 47 receive score~2 (vague) and 37 score~3 (insufficient information).

Test-coverage adequacy shows a higher baseline: 382 instances (70.6\%) receive score~0 and 60 (11.1\%) score~1.
A total of 99 instances (18.3\%) are excluded for inadequate test coverage, of which 43 score~2 and 56 score~3.
Fifty-five instances (10.2\%) are excluded for \emph{both} reasons simultaneously, yielding 134 total exclusions (24.8\%) and 407 verified benchmark instances (75.2\%).

\subsection{Task Characteristics}

Figure~\ref{fig:task-characteristics} summarises the composition of the 407 verified instances by estimated resolution time and issue type.
Difficulty is broadly distributed: 138 instances (33.9\%) are estimated to take under 15 minutes, 139 (34.2\%) 15 minutes to one hour, 78 (19.2\%) one to four hours, and 52 (12.8\%) more than four hours.
By issue type, 180 instances (44.2\%) are Bug Fixes, 146 (35.9\%) are New Features, and 81 (19.9\%) are Feature Optimizations.

\section{Benchmark Comparison}
\label{sec:benchmarks-comparison}

\subsection{Comparison with Existing Benchmarks}
\label{subsec:benchmark-comparison}

Table~\ref{tab:benchmark-comparison} situates {\benchmark} among existing issue-resolution benchmarks across four dimensions: domain coverage, verification, difficulty stratification, and patch complexity. Most prior benchmarks focus on library-style repositories, often in a single language or a small set of languages. Their gold patches are typically small, averaging between 1.2 and 4.9 modified files and between 14 and 163 changed lines. In contrast, {\benchmark} targets production mobile apps and exhibits substantially larger fixes, averaging 12.9 modified files and 334.6 changed lines.

\noindent\textbf{Patch Complexity.}
Figure~\ref{fig:patch-complexity} compares per-repository median patch size across five benchmarks using three metrics: files modified, hunks, and changed lines. SWE-PolyBench~\citep{2025swepolybench}, SWE-bench Multimodal~\citep{yang2024swebenchmultimodal}, and Multi-SWE-bench~\citep{multi2025swebench} occupy a compact low-to-moderate range, with median file counts of 2.6, 3.6, and 3.0, and median changed lines of 54, 128, and 73, respectively.
Rust-SWE-bench~\citep{2025rustswebench} is an outlier on the files axis (median 10.4 files) owing to the broad cross-module scope of typical Rust fixes, yet its median changed lines (92) remain well below those of {\benchmark}.

{\benchmark} presents the largest median hunk count (27.5) and changed line count (342.6) across all five benchmarks. Compared with SWE-PolyBench, the closest match on repository count, {\benchmark} fixes are 3.1$\times$ larger in files, 3.9$\times$ in hunks, and 6.3$\times$ in changed lines. Relative to Multi-SWE-bench, the ratios are 2.7$\times$, 3.6$\times$, and 4.7$\times$, whereas relative to SWE-bench Multimodal, they are 2.3$\times$, 3.5$\times$, and 2.7$\times$.
While Rust-SWE-bench has a slightly higher median file count than {\benchmark} (10.4 vs.\ 8.1), {\benchmark} exceeds it by 2.7$\times$ in hunks and 3.7$\times$ in changed lines, indicating that mobile app fixes involve more densely modified changes rather than touching more files.
This pattern reflects the coordinated nature of mobile app fixes, where a single logical change often requires updates across source code, resources, and build configuration files.

\noindent\textbf{Design Dimensions and Scale.}
{\benchmark} contains 407 manually verified instances from 19 repositories across four languages, comparable in scale to SWE-PolyBench Verified~\citep{2025swepolybench} (382 instances, 20 repositories) and roughly four-fifths of SWE-bench Verified~\citep{openai2024swebenchverified} (500 instances, 12 repositories), while targeting a distinct application domain. Beyond scale, it differs from prior benchmarks in three design aspects. First, it covers production mobile apps spanning three frameworks (Android Native, React Native, and Flutter) across four programming languages, a domain absent from all existing benchmarks. Second, it combines human verification with difficulty stratification, enabling controlled evaluation across task complexity levels. Third, it explicitly captures multi-artifact fixes, where resolving an issue requires coordinated changes across heterogeneous project components rather than isolated source edits. Issue descriptions are also longer on average (308 tokens), reflecting the richer contextual information typical of mobile bug reports, such as reproduction steps, device configurations, and platform versions.



\subsection{Conceptual Differences from Backend Benchmarks}

Existing benchmarks, such as SWE-bench~\citep{jimenez2024swebench} and its multimodal~\citep{yang2024swebenchmultimodal} and multilingual variants~\citep{multi2025swebench, 2025swepolybench, kabir2025swebenchmultilingual}, primarily measure an LLM's capacity to resolve isolated logic errors within generic programming environments. These benchmarks often evaluate library-centric or backend repositories where control flow is direct and execution environments are relatively straightforward (e.g., standard Python or JavaScript runtimes).

\textbf{Execution and Lifecycle:} Mobile application frameworks fundamentally change this control flow. The operating system dictates the lifecycle of an application, instantiating UI components (such as Activities in Android or Widgets in Flutter) and maintaining complex asynchronous states. MobileDev-Bench tests an LLM's ability to reason over these implicit callbacks and event-driven architectures that are entirely absent in CLI or library benchmarks.

\textbf{Artifact Diversity:} Furthermore, a core distinction of MobileDev-Bench is its emphasis on multi-artifact coordination. While conventional benchmarks evaluate patches applied strictly to source code files, mobile fixes routinely demand coordinated changes across declarative user interface definitions (e.g., XML layouts), localized string resources, manifestation configurations, and heavy build scripts (Gradle). Models that excel at pure algorithmic synthesis often falter when required to maintain synchronization between a deeply nested source logic and external framework resources.

\textbf{Scale of Modifications:} As detailed in the comparison in Figure ~\ref{fig:patch-complexity}, the sheer scale of patches in MobileDev-Bench far exceeds those typical of SWE-bench paradigms. The necessity to touch multiple files reflects the cross-cutting operational reality of mobile deployments, setting a new bar for evaluation environments emphasizing integration robustness.

\begin{table}[t]
\centering
\footnotesize
\setlength{\tabcolsep}{3pt}
\caption{Framework-wise instance distribution and code change characteristics in MobileDev-Bench}
\label{tab:language-distribution}
\begin{tabular}{lrrr}
\toprule
\textbf{Metric} & \textbf{Android Native} & \textbf{Flutter} & \textbf{React-Native} \\
\midrule
\multicolumn{4}{l}{\textit{Instance distribution}} \\
\cmidrule(lr){1-4}
Repositories & 13 & 2 & 4 \\
Instances & 328 & 63 & 16 \\
\midrule
\multicolumn{4}{l}{\textit{Fix patch statistics}} \\
\cmidrule(lr){1-4}
Avg. \# files & 14.0 & 7.6 & 9.5 \\
Avg. \# hunks & 28.8 & 18.3 & 36.2 \\
Lines added & 222.3 & 204.6 & 314.8 \\
Lines removed & 119.6 & 50.6 & 181.8 \\
\midrule
\multicolumn{4}{l}{\textit{Test patch statistics}} \\
\cmidrule(lr){1-4}
Avg. \# files & 6.1 & 4.0 & 3.4 \\
Avg. \# hunks & 15.9 & 13.9 & 15.2 \\
Lines added & 141.8 & 234.3 & 102.7 \\
Lines removed & 50.7 & 43.5 & 70.1 \\
\bottomrule
\end{tabular}
\end{table}

\section{Dataset Analysis}
\label{sec:dataset-analysis}

We provide additional quantitative analyses of the {\benchmark} dataset that complement the summary statistics in Section~\ref{sec:characterization}.

\subsection{Framework-wise Distribution and Code Change Characteristics}

Table~\ref{tab:language-distribution} reports per-framework instance counts and patch statistics.
Android Native dominates the dataset (328 instances, 80.6\%), reflecting the larger number of mature open-source Android repositories.
Flutter contributes 63 instances (15.5\%) from two high-activity repositories, and React Native contributes 16 instances (3.9\%) from four repositories.
The fix patch statistics reveal that Android Native instances involve substantially more code changes on average (14.0 modified files, 28.8 hunks) than Flutter (7.6 files, 18.3 hunks), reflecting the greater modularisation typical of large Android codebases.
Notably, Flutter test patches add considerably more lines on average (234.3) than their fix counterparts (204.6), indicating that Flutter contributors tend to write comprehensive test suites alongside their fixes.

\subsection{Distribution of Artifact Type Combinations}

Table~\ref{tab:artifact-combinations} reports the distribution of artifact type combinations present in fix patches.
The majority of instances (58.5\%) modify only source files, and 59.0\% are single-artifact overall, indicating that the primary challenge is code comprehension and logic repair.
However, 41.0\% of instances require coordinated changes across multiple artifact types, testing a model's ability to reason about the full mobile-development toolchain.
The most common multi-artifact pattern is Source + Build (8.1\%), where fixes also modify Gradle build scripts, dependency version catalogues, or build configuration files.
Android Native accounts for most multi-artifact instances: patterns combining source with resources, internationalisation string files, or build scripts are exclusive to or heavily concentrated in Android projects, whereas Flutter instances are predominantly source-only (74.6\%).


\begin{table*}[t]
\centering
\footnotesize
\setlength{\tabcolsep}{3.5pt}
\caption{Distribution of artifact type combinations in MobileDev-Bench. 41.0\% of instances require coordinated changes across multiple artifact types.}
\label{tab:artifact-combinations}
\begin{tabular}{lrr|rrr}
\toprule
& \multicolumn{2}{c|}{\textbf{Overall}} & \multicolumn{3}{c}{\textbf{\% by Framework}} \\
\cmidrule(lr){2-3}\cmidrule(lr){4-6}
\textbf{Combination} & \textbf{\#} & \textbf{\%} & \textbf{Android Native} & \textbf{Flutter} & \textbf{React Native} \\
\midrule
\multicolumn{6}{l}{\textit{Single Artifact Type}} \\
Src & 238 & 58.5 & 55.8 & 74.6 & 50.0 \\
Res & 1 & 0.3 & 0.3 & 0.0 & 0.0 \\
Build & 1 & 0.2 & 0.3 & 0.0 & 0.0 \\
\textit{Subtotal} & 240 & 59.0 & 56.4 & 74.6 & 50.0 \\
\midrule
\multicolumn{6}{l}{\textit{Two Artifact Types}} \\
Src + Build & 33 & 8.1 & 10.0 & 0.0 & 0.0 \\
Src + Res & 15 & 3.7 & 4.6 & 0.0 & 0.0 \\
Src + i18n & 15 & 3.7 & 0.0 & 20.6 & 12.5 \\
Src + Docs & 13 & 3.2 & 4.0 & 0.0 & 0.0 \\
Src + Other & 4 & 0.9 & 0.9 & 0.0 & 6.3 \\
Other & 2 & 0.5 & 0.3 & 1.6 & 0.0 \\
\textit{Subtotal} & 82 & 20.1 & 19.8 & 22.2 & 18.8 \\
\midrule
\multicolumn{6}{l}{\textit{Three or More Artifact Types}} \\
Src + Res + i18n & 27 & 6.7 & 8.2 & 0.0 & 0.0 \\
Src + Res + i18n + Build & 11 & 2.7 & 3.4 & 0.0 & 0.0 \\
Src + Build + Docs & 7 & 1.7 & 1.8 & 1.6 & 0.0 \\
Src + Manifest + Config & 5 & 1.2 & 0.0 & 1.6 & 25.0 \\
Src + Res + i18n + Docs & 5 & 1.2 & 1.5 & 0.0 & 0.0 \\
Other & 30 & 7.4 & 8.9 & 0.0 & 6.2 \\
\textit{Subtotal} & 85 & 20.9 & 23.8 & 3.2 & 31.2 \\
\midrule
\textbf{Total (Single)} & \textbf{240} & \textbf{59.0} & \textbf{56.4} & \textbf{74.6} & \textbf{50.0} \\
\textbf{Total (Multi)} & \textbf{167} & \textbf{41.0} & \textbf{43.6} & \textbf{25.4} & \textbf{50.0} \\
\bottomrule
\end{tabular}
\end{table*}

\subsection{Background Knowledge Required by Framework}
\label{subsec:background-knowledge}

\begin{figure}[t]
  \centering
  \includegraphics[width=\columnwidth]{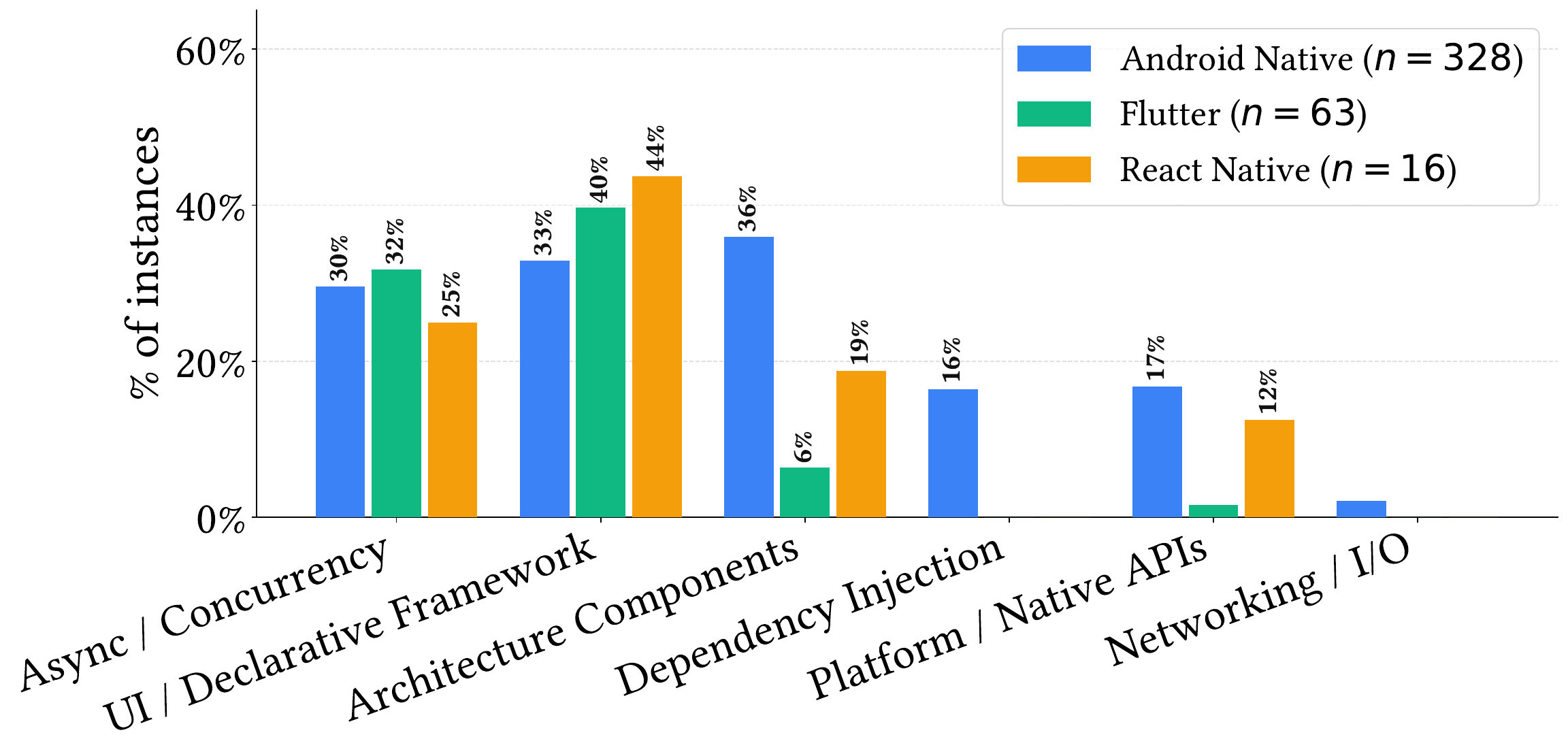}
  \caption{Percentage of instances per framework whose fix patches involve
    each technical knowledge domain, inferred from import statements,
    annotations, and identifiers in the added lines of fix patches.
    An instance may span multiple domains.}
  \label{fig:background-knowledge}
\end{figure}

Figure~\ref{fig:background-knowledge} characterises the technical knowledge domains implicitly required by each framework's instances, inferred from import statements, annotations, and identifiers in the added lines of fix patches.
The six domains span the main layers of modern mobile engineering: asynchronous programming, declarative UI frameworks, architecture components, dependency injection, platform and native APIs, and networking.

For Android Native, Architecture Components are the most frequently required domain (36.0\%), reflecting widespread use of \texttt{ViewModel}, \texttt{Room}, \texttt{WorkManager}, and Navigation across the benchmark repositories.
Declarative UI knowledge (Jetpack Compose) is needed in 32.9\% of Android Native instances, Asynchronous/Concurrency patterns (Kotlin coroutines, \texttt{Flow}) in 29.6\%, Platform SDK APIs in 16.8\%, and Dependency Injection via Hilt or Dagger in 16.5\%.
Together, these figures confirm that Android Native instances stress a broad stack of modern Android engineering beyond core Kotlin syntax.

Flutter instances primarily demand Widget and UI framework knowledge (39.7\%) alongside asynchronous Dart patterns (31.7\%), consistent with Flutter's widget-centric programming model.
Architecture and state-management patterns (BLoC, Provider) are needed in only 6.3\% of Flutter instances, reflecting the dataset's focus on high-activity bug-fix PRs rather than large architectural changes.

React Native percentages are indicative only ($n\!=\!16$): UI framework patterns appear in 43.8\% of instances and Async/Concurrency in 25.0\%, with Architecture Components in 18.8\% and Platform/Native APIs in 12.5\%; Dependency Injection and Networking domains register zero, plausibly reflecting these repositories' reliance on abstracted, library-level patterns not captured by source-level identifiers alone.

\section{Experimental Setup}
\label{sec:appendix_experimental_setup}

This section provides detailed information about the experimental framework, model selection, and evaluation metrics used to assess automated program repair performance on MobileDev-Bench.

\subsection{Evaluation Framework}

We evaluate MobileDev-Bench using the localization and repair pipeline from Agentless~\citep{xia2024agentless}, a state-of-the-art autonomous code repair framework that has demonstrated strong performance on Python-centric benchmarks. Agentless decomposes the repair task into two sequential phases:

\begin{enumerate}
    \item \textbf{Fault Localization:} Given an issue description and repository context, the model identifies potentially buggy files and functions. This phase employs retrieval-augmented generation to narrow the search space from the entire codebase to a manageable set of candidate locations.

    \item \textbf{Patch Generation:} The model generates a patch as a unified diff format targeting the localized code elements. The generated patch must be syntactically valid and applicable to the base commit.
\end{enumerate}

This two-phase decomposition enables us to separately measure localization capability (can the model identify where to fix?) and repair capability (can the model generate the correct fix?), providing diagnostic insights into failure modes.

We validate generated patches using a Docker-based test harness. For each task instance, we: (1) clone the repository at the base commit, (2) apply the model-generated patch, (3) execute the repository test suite in an isolated Docker container, and (4) compare test outcomes against the ground truth. A patch is considered successful if it passes all tests that pass with the developer's ground truth patch, including previously failing tests that should now pass.

This evaluation approach mirrors the workflow used in SWE-bench~\citep{jimenez2024swebench} and Multi-SWE-bench~\citep{multi2025swebench}, enabling direct comparison with existing benchmarks while avoiding the compounding errors and non-determinism of full agentic systems that include web browsing, terminal access, and multi-turn interactions.

\subsection{Adaptation for Mobile Development Languages}
\label{app:lang-adaptation}

The original Agentless implementation relies on Python's Abstract Syntax Tree (AST) module for parsing and extracting code structures. To support the diverse programming languages in MobileDev-Bench (Java, Kotlin, TypeScript, and Dart), we incorporate tree-sitter\footnote{\url{https://pypi.org/project/tree-sitter-language-pack/}}, a parsing library that provides unified syntax tree generation across multiple languages.

Similar to the MagentLess\footnote{\url{https://github.com/multi-swe-bench/MagentLess}} adaptation for Multi-SWE-bench, we replace Python-specific parsing calls with tree-sitter-based parsing while maintaining the same localization and patch generation algorithms. This ensures that:

\begin{itemize}
    \item Code structure extraction (classes, methods, functions) works consistently across Android (Java/Kotlin), Flutter (Dart), and React Native (TypeScript/JavaScript) codebases.
    \item Function-level localization operates with the same granularity across all four languages.
    \item The core Agentless algorithm remains unchanged, ensuring fair comparison with results on Python-based benchmarks.
\end{itemize}

For non-code artifacts (XML layouts, JSON i18n files, Gradle build scripts), which lack traditional function boundaries, we perform file-level localization only.

\subsection{Model Selection}
\label{sec:model-selection}

We evaluate four frontier language models: Claude Sonnet 4.5~\citep{anthropic2025claude} (Anthropic's proprietary model with 1M token context window), GPT-5.2~\citep{openai2025gpt52} (OpenAI's frontier model with enhanced reasoning capabilities), Gemini 2.5 Flash~\citep{comanici2025gemini25pushingfrontier} (Google's fast-inference model with native multimodal support), and Qwen-3-Coder~\citep{qwen2025qwen3coder} (an open-weight model from Alibaba Cloud, fine-tuned for code tasks).

All models are accessed via OpenRouter\footnote{\url{https://openrouter.ai/}}.

\textbf{Agentless setting.} We use temperature 0.0 for fault localization. For patch generation, we generate one greedy sample (temperature 0.0) and multiple diverse samples (temperature 0.8), then rank and select the best patch. To ensure fair comparison, we disable extended reasoning modes: \texttt{reasoning\_effort=low} for GPT-5.2, no extended thinking for Claude Sonnet 4.5, and default settings for Gemini 2.5 Flash and Qwen-3-Coder.

\textbf{Oracle setting.} The fault localization step is bypassed; models are given the exact set of files modified in the ground truth patch. Patch generation uses a single greedy sample (temperature 0.0). No reasoning mode overrides are applied.


\begin{table*}[t]
\centering
\small
\setlength{\tabcolsep}{4pt}
\caption{Resolution rates by ground truth file count under the Automated and Oracle retrieval settings. Rate is resolved / evaluated instances (instances without a submitted prediction excluded in Automated retrieval).}
\label{tab:resolution-rate}
\begin{tabular}{lrcccccccc}
\toprule
& & \multicolumn{4}{c}{\textbf{Resolved}} & \multicolumn{4}{c}{\textbf{Rate (\%)}} \\
\cmidrule(lr){3-6} \cmidrule(lr){7-10}
\textbf{\#Files} & \textbf{Tasks} & \textbf{C.~Sonnet 4.5} & \textbf{Qwen3} & \textbf{GPT~5.2} & \textbf{G.~2.5} & \textbf{C.~Sonnet 4.5} & \textbf{Qwen3} & \textbf{GPT~5.2} & \textbf{G.~2.5} \\
\midrule
\multicolumn{10}{l}{\textit{Automated Retrieval}} \\
1 file       &  71 & 11 &  9 & 11 &  9 &  15.5 &  12.7 &  15.5 &  12.7 \\
2--3 files   &  89 &  4 &  2 &  2 &  2 &   4.5 &   2.2 &   2.2 &   2.2 \\
4--5 files   &  52 &  2 &  2 &  0 &  2 &   3.8 &   3.8 &   0.0 &   3.8 \\
6--10 files  &  70 &  0 &  0 &  0 &  0 &   0.0 &   0.0 &   0.0 &   0.0 \\
11+ files    & 125 &  0 &  0 &  0 &  0 &   0.0 &   0.0 &   0.0 &   0.0 \\
\cmidrule(lr){1-10}
Overall      & 407 & 17 & 13 & 13 & 13 &  4.23 &  3.33 &  3.23 &  3.24 \\
\midrule
\multicolumn{10}{l}{\textit{Oracle Retrieval}} \\
1 file       &  71 & 18 &  9 & 14 &  7 &  25.4 &  12.7 &  19.7 &   9.9 \\
2--3 files   &  89 &  3 &  1 &  1 &  0 &   3.4 &   1.1 &   1.1 &   0.0 \\
4--5 files   &  52 &  2 &  0 &  1 &  1 &   3.8 &   0.0 &   1.9 &   1.9 \\
6--10 files  &  70 &  0 &  0 &  2 &  0 &   0.0 &   0.0 &   3.0 &   0.0 \\
11+ files    & 125 &  0 &  0 &  0 &  0 &   0.0 &   0.0 &   0.0 &   0.0 \\
\cmidrule(lr){1-10}
Overall      & 407 & 23 & 10 & 18 &  8 &  5.69 &  2.57 &  4.53 &  1.98 \\
\bottomrule
\end{tabular}
\end{table*}

\subsection{Evaluation Metrics}
\label{app:appendix-metrics}

Following prior work on automated program repair~\citep{xia2024agentless, 2025swepolybench}, we report two complementary families of metrics that separately assess end-to-end repair capability and fault localization accuracy.

\subsubsection{Resolution Rate}

Let $T$ denote the total number of tasks in the benchmark. For each task $t$, we define an indicator function $\mathbf{1}(\mathrm{Pass}_t)$ that equals 1 if the model-generated patch passes all tests when applied to the base commit, and 0 otherwise. We compute Resolution Rate as:

\[
\mathrm{Resolution\ Rate} = \frac{1}{T} \sum_{t=1}^{T} \mathbf{1}(\mathrm{Pass}_t)
\]

This metric measures end-to-end repair capability at the task level. A high Resolution Rate indicates that the model successfully localizes the bug, generates a syntactically correct patch, and produces semantically correct fixes that satisfy all test requirements.

Resolution Rate is the primary metric for comparing model performance, as it directly measures the practical utility of automated repair systems. However, it conflates localization and generation capabilities, providing limited diagnostic insight when rates are low.

\subsubsection{Retrieval Metrics: Precision, Recall, and F1}

To isolate fault localization performance from patch generation quality, we compute file-level retrieval metrics. For each task $t$, let $F^{GT}_t$ denote the set of ground truth modified files (extracted from the developer's patch) and let $F^{Pred}_t$ denote the set of files modified by the model-generated patch.

We compute per-task retrieval metrics as:

\begin{align*}
\mathrm{Recall}_t &= \frac{|F^{GT}_t \cap F^{Pred}_t|}{|F^{GT}_t|} \\
\mathrm{Precision}_t &= \frac{|F^{GT}_t \cap F^{Pred}_t|}{|F^{Pred}_t|} \\
\mathrm{F1}_t &= \frac{2 \cdot \mathrm{Precision}_t \cdot \mathrm{Recall}_t}{\mathrm{Precision}_t + \mathrm{Recall}_t}
\end{align*}

We then report macro-averaged metrics across all $T$ tasks:

\begin{align*}
\mathrm{Recall} &= \frac{1}{T} \sum_{t=1}^{T} \mathrm{Recall}_t \\
\mathrm{Precision} &= \frac{1}{T} \sum_{t=1}^{T} \mathrm{Precision}_t \\
\mathrm{F1} &= \frac{1}{T} \sum_{t=1}^{T} \mathrm{F1}_t
\end{align*}

These metrics measure fault localization performance independently of patch correctness:

\begin{itemize}
    \item \textbf{Recall} captures the fraction of ground truth files that the model correctly identifies as requiring modification. Low recall indicates the model misses many necessary modification sites.

    \item \textbf{Precision} measures the fraction of predicted files that are actually relevant (i.e., appear in the ground truth). High precision indicates the model avoids editing irrelevant files, while low precision suggests over-editing or incorrect localization.

    \item \textbf{F1} provides a harmonic mean balancing both aspects, penalizing models that achieve high recall through indiscriminate file editing or high precision through overly conservative predictions.
\end{itemize}

Together with Resolution Rate, these metrics distinguish localization capability from repair capability. High recall with low Resolution Rate suggests that the model identifies the correct files but fails to generate correct fixes. Conversely, low recall limits achievable Resolution Rate, since a model cannot repair files it fails to localize. High precision indicates the model avoids over-editing irrelevant files, which is important for minimizing unintended side effects in production deployments.

\subsection{Evaluation Infrastructure}
\label{sec:evaluation-infrastructure}

Each task instance is executed in an isolated Docker container with repository-specific dependencies installed from the original project's build configuration (Gradle for Android, pub for Flutter, npm/yarn for React Native).

Test execution timeouts are set to 30 minutes per task to accommodate large test suites in production repositories. Tasks that exceed this limit are marked as timeouts and counted as failures.

\section{Evaluation Results}
\label{sec:appendix_evaluation_results}

This section presents comprehensive evaluation results for four frontier language models on MobileDev-Bench. We analyze resolution rates, fault localization performance, and the relationship between task characteristics and repair success.

\subsection{Resolution Rate Analysis}

Table~\ref{tab:resolution-rate} shows resolution rates for both evaluation settings on 407 verified tasks (rate = resolved / evaluated, excluding empty-patch predictions from the denominator). Under the Automated retrieval setting, rates range from 3.23\% to 4.23\% (13--17 resolved per model). Under the Oracle retrieval setting, rates range from 1.98\% (Gemini 2.5 Flash, 8 resolved) to 5.69\% (Claude Sonnet 4.5, 23 resolved). Tasks in MobileDev-Bench require modifying an average of 12.9 files (median: 5), reflecting the multi-file nature of mobile development.

\vspace{10pt}
\noindent\textbf{Multi-File Modifications Show Lower Resolution Rates.}

\begin{figure}[t]
\centering
\includegraphics[width=0.48\textwidth]{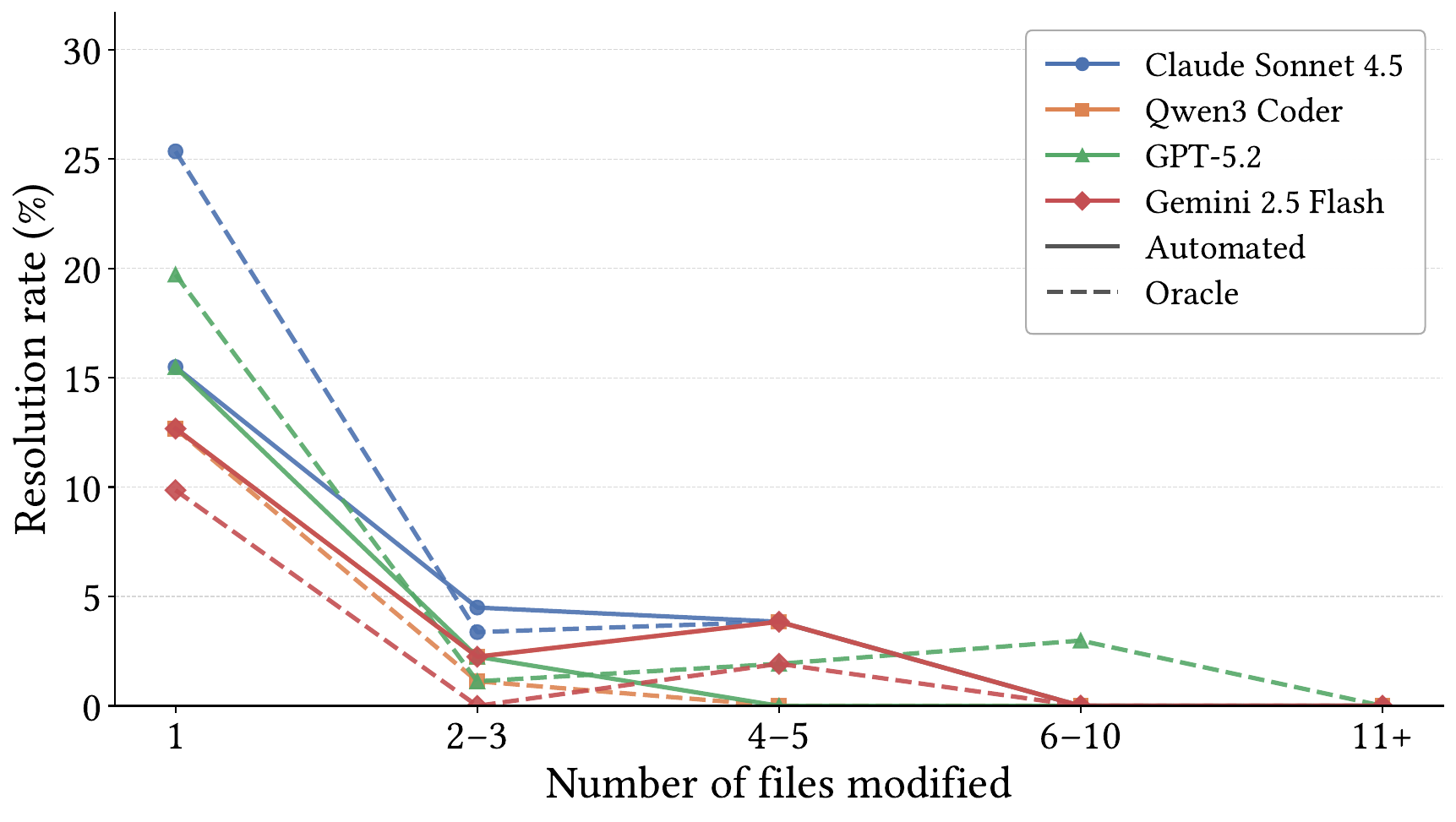}
\caption{Resolution rates by number of files modified, for Automated (solid lines) and Oracle (dashed lines) retrieval settings. Single-file tasks achieve 12.7--15.5\% under Automated and up to 25.4\% under Oracle, dropping to near zero for 6+ files.}
\label{fig:resolution-filecount}
\end{figure}

Figure~\ref{fig:resolution-filecount} reveals a dramatic degradation in resolution rates as the number of required file modifications increases. Under the Automated retrieval setting, single-file tasks (71 tasks) achieve 12.7--15.5\% resolution rates, substantially higher than the 3.23--4.23\% overall rates. Under the Oracle retrieval setting, single-file rates rise further to 9.9--25.4\%, reflecting that localization is the dominant bottleneck on simpler tasks. However, success rapidly diminishes with task complexity under both settings.

Tasks requiring 2--3 file modifications show a steep drop to 2.2--4.5\% under Automated retrieval and 0.0--3.4\% under Oracle retrieval. Mid-complexity tasks requiring 4--10 files prove nearly unsolvable (0.0--3.8\% Automated retrieval; 0.0--3.8\% Oracle retrieval). This difficulty cliff highlights models' fundamental limitations in coordinating changes across multiple files.

Tasks requiring 11+ files (125 tasks, 30.7\% of the benchmark) yield 0.0\% resolution across all models under both Automated and Oracle retrieval settings.

Across all models, only 28 unique instances were successfully resolved under Automated retrieval, accumulating to 56 total resolutions due to overlap. The remaining 379 tasks (93.1\%) were not resolved by any model under the Automated retrieval setting, underscoring the fundamental challenge of multi-file reasoning in mobile development contexts.

\vspace{15pt}
\noindent\textbf{Larger Patches Correlate with Lower Success Rates.}

\begin{figure}[t]
\centering
\includegraphics[width=0.48\textwidth]{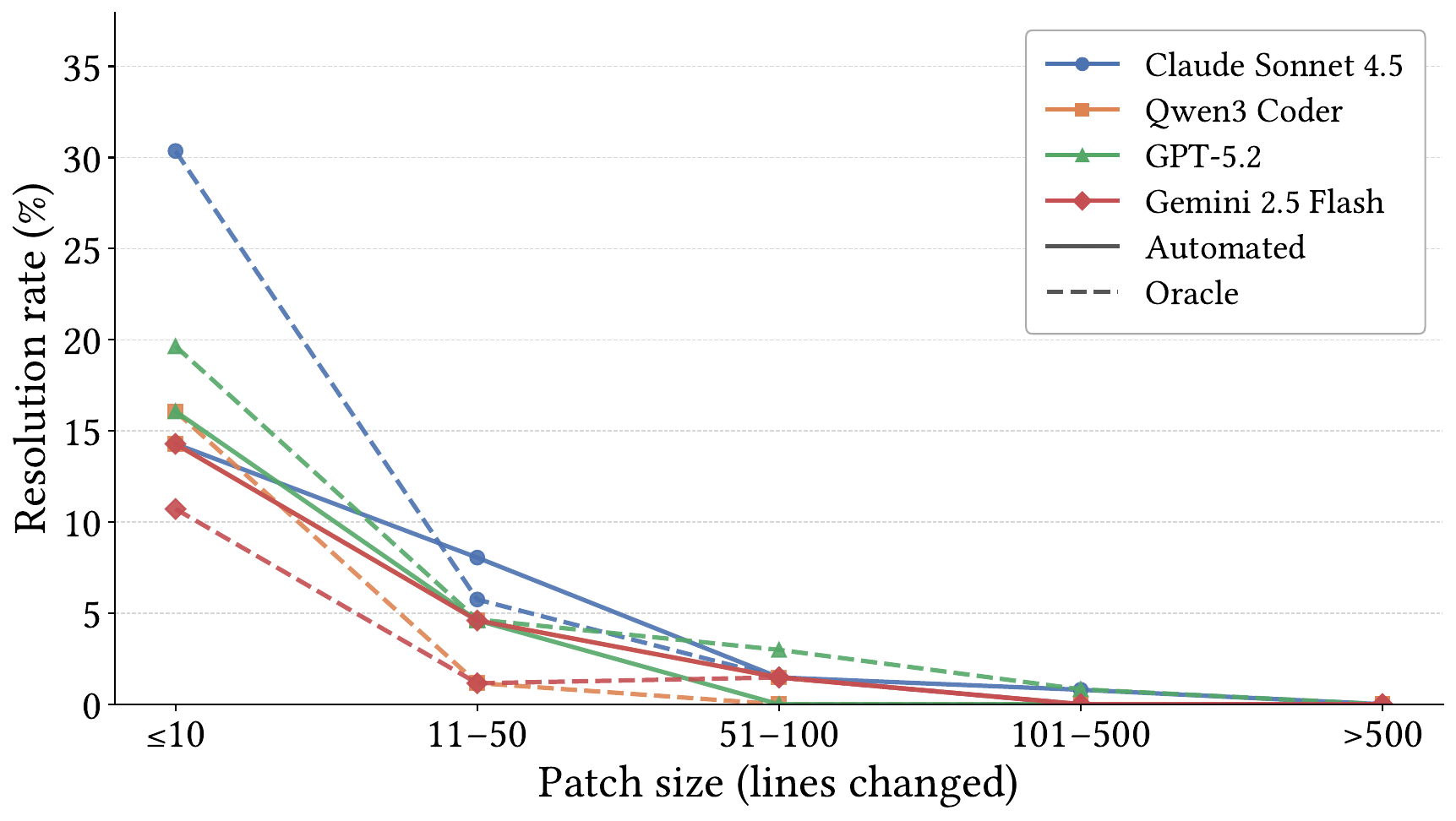}
\caption{Resolution rates by patch size, for Automated (solid lines) and Oracle (dashed lines) retrieval. Tiny patches ($\leq$10 lines) achieve 14.3--16.1\% under Automated and 10.7--30.4\% under Oracle, dropping to near zero for patches $>$100 lines.}
\label{fig:resolution-patchsize}
\end{figure}

Patch complexity, measured by lines changed in the ground truth fix, strongly predicts resolution success (Figure~\ref{fig:resolution-patchsize}). Under Automated retrieval, tiny patches ($\leq$10 lines) achieve 14.3--16.1\% resolution rates, while very large patches ($>$500 lines) yield 0.0\% across all models. Under Oracle retrieval, rates for $\leq$10-line patches reach 10.7--30.4\%, and $>$500-line patches remain at 0.0\%. This inverse relationship reveals that models struggle not only with multi-file coordination but also with generating extensive code changes within identified locations.

Resolved instances exhibit dramatically smaller patches than unresolved tasks. The median patch size for resolved instances is 7.0 lines versus 106.0 lines for unresolved instances (15.1x difference). Mean patch sizes show a similar pattern: 16.4 lines for resolved versus 358.1 lines for unresolved, a 95\% reduction. This size difference indicates that successful repairs target localized, focused changes rather than broad refactorings.

\begin{figure}[t]
\centering
\includegraphics[width=0.48\textwidth]{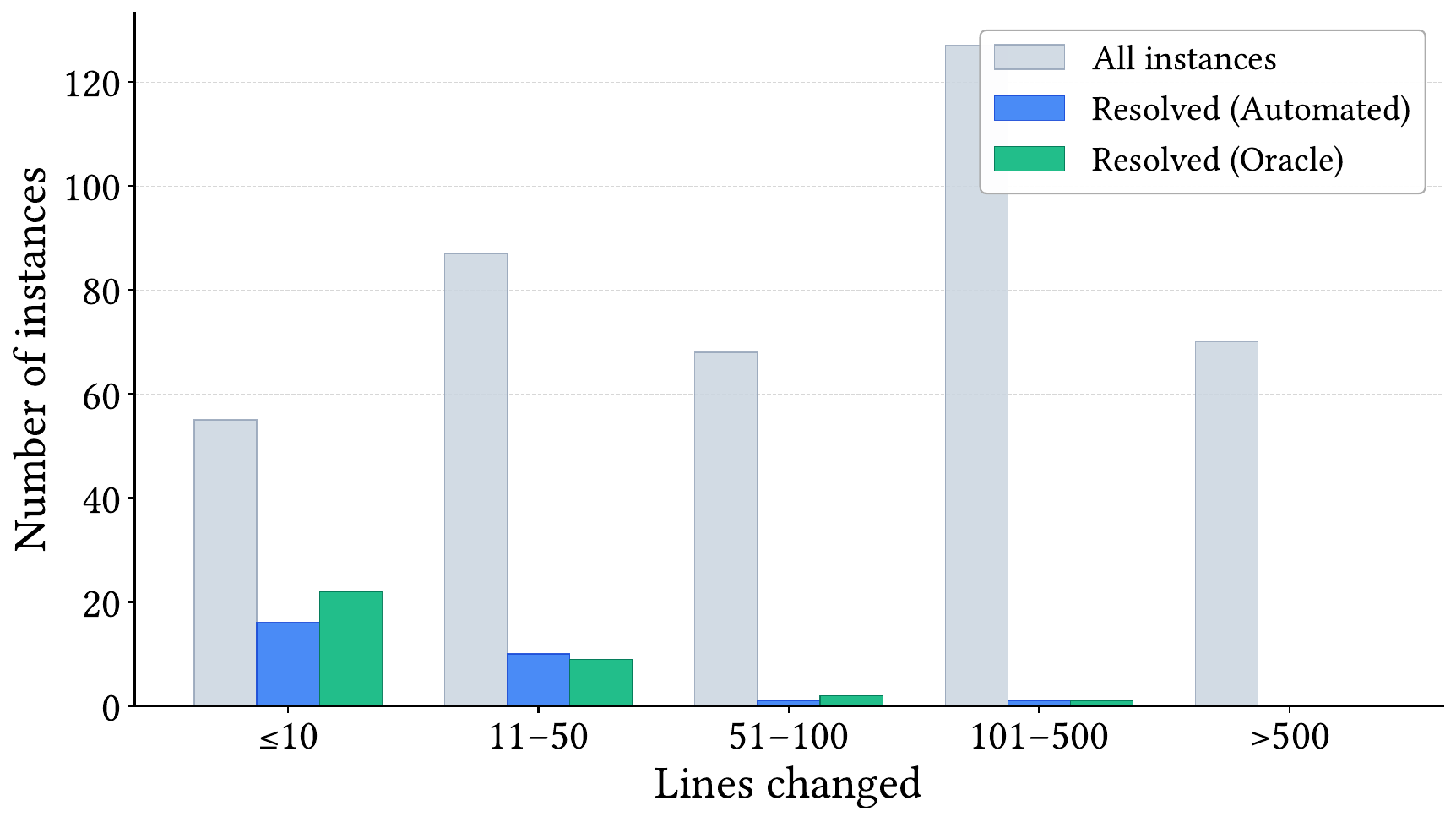}
\caption{Patch size distribution for all instances versus resolved instances. Resolved instances skew heavily toward patches $\leq$ 50 lines (92.9\% vs. 35.1\%).}
\label{fig:patchsize-distribution}
\end{figure}

Patch size distributions further emphasize this disparity (Figure~\ref{fig:patchsize-distribution}). Small patches ($\leq$50 lines) comprise 92.9\% of resolutions but only 35.1\% of the benchmark, a 2.6x overrepresentation. Conversely, patches $>$50 lines represent 64.9\% of the benchmark but only 7.1\% of resolutions. Tiny patches ($\leq$10 lines) show the most extreme skew: 60.7\% of resolutions versus 13.8\% of benchmark (4.4x overrepresentation). Resolution rates collapse at medium patch sizes (51--100 lines: 0.0--1.5\%) and remain near zero for larger patches (101--500 lines: 0.0--0.8\%; $>$500 lines: 0.0\%).

This finding complements the file count analysis, revealing two orthogonal dimensions of repair difficulty: spatial distribution (how many files) and change magnitude (how many lines). While file count measures coordination complexity, patch size measures generation complexity. Large patches require models to reason about extensive code interactions, maintain consistency across many modified lines, and ensure all changes compile and pass tests. Current models excel at small, surgical fixes but fail when extensive code generation is required.

\subsection{Characteristics of Resolved Instances}

Table~\ref{tab:resolution-rate-detailed} analyzes how resolved instances differ from the overall benchmark distribution across three dimensions: artifact type diversity, file count distribution, and framework. This analysis reveals structural factors that influence repair success.


\begin{table}[t]
\centering
\footnotesize
\setlength{\tabcolsep}{4pt}
\caption{Resolution rate breakdown by task characteristics (407 tasks). Resolved column shows the share of uniquely resolved instances (28 under Automated, 34 under Oracle) in each category vs.\ the overall benchmark share. Ratio = Resolved\% / Benchmark\%.}
\label{tab:resolution-rate-detailed}
\begin{tabular}{lrcccc}
\toprule
& & \multicolumn{2}{c}{\textbf{Automated Retrieval}} & \multicolumn{2}{c}{\textbf{Oracle Retrieval}} \\
\cmidrule(lr){3-4}\cmidrule(lr){5-6}
\textbf{Category} & \textbf{Bench.} & \textbf{Resolved} & \textbf{Ratio} & \textbf{Resolved} & \textbf{Ratio} \\
\midrule
\multicolumn{6}{l}{\textit{Artifact Type Diversity}} \\
Single-artifact &  59.0\% & 96.4\% & 1.63x & 97.1\% & 1.65x \\
Multi-artifact  &  41.0\% &  3.6\% & 0.09x &  2.9\% & 0.07x \\
\midrule
\multicolumn{6}{l}{\textit{File Count Distribution}} \\
1 file          &  17.4\% & 75.0\% & 4.31x & 76.5\% & 4.40x \\
2--3 files      &  21.9\% & 17.9\% & 0.82x & 11.7\% & 0.53x \\
4--5 files      &  12.8\% &  7.1\% & 0.55x &  5.9\% & 0.46x \\
6--10 files     &  17.2\% &  0.0\% & 0.00x &  5.9\% & 0.34x \\
11+ files       &  30.7\% &  0.0\% & 0.00x &  0.0\% & 0.00x \\
\midrule
\multicolumn{6}{l}{\textit{Framework Distribution}} \\
Android Native  &  80.6\% & 67.9\% & 0.84x & 82.4\% & 1.02x \\
Flutter         &  15.5\% & 32.1\% & 2.07x & 14.7\% & 0.95x \\
React Native    &   3.9\% &  0.0\% & 0.00x &  2.9\% & 0.74x \\
\bottomrule
\end{tabular}
\vspace{-18pt}
\end{table}


\vspace{5pt}
\noindent\textbf{Artifact Type Diversity.} We classify tasks as single-artifact (all files of one type) or multi-artifact (spanning source code, resources, build files, manifests). While 59.0\% of benchmark tasks are single-artifact, 96.4\% of automated retrieval resolutions come from single-artifact tasks (1.63x overrepresentation). Multi-artifact tasks show severe underrepresentation (3.6\% of resolutions vs. 41.0\% of benchmark, 0.09x), highlighting difficulty coordinating changes across heterogeneous file types.

\vspace{5pt}
\noindent\textbf{File Count Distribution.}
Single-file tasks dominate automated retrieval resolutions (75.0\% of resolutions vs. 17.4\% of benchmark, 4.31x overrepresentation), while multi-file tasks show severe underrepresentation. Tasks requiring 6--10 files exhibit 0.0\% of automated retrieval resolutions versus 17.2\% of benchmark (0.00x ratio). The 11+ file category also yields 0.0\% resolutions (0.00x ratio), despite comprising 30.7\% of the benchmark.

\vspace{5pt}
\noindent\textbf{Framework Distribution.}
\label{sec:appendix_results_framework}
Framework distribution among resolved tasks broadly follows benchmark composition but with notable deviations (Table~\ref{tab:resolution-rate-detailed}). Android Native tasks are slightly underrepresented (67.9\% of Automated retrieval resolutions vs. 80.6\% of benchmark, 0.84x), while Flutter tasks are overrepresented at 32.1\% of resolutions despite comprising only 15.5\% of the benchmark (2.07x), likely reflecting the structural characteristics of Flutter repositories, which tend toward single-artifact changes with strong semantic signals in issue descriptions. React Native tasks see complete failure under Automated retrieval (0.0\% vs. 3.9\%), mirroring patterns seen across complex multi-layer architectures where cross-file dependency tracing breaks down. These differences reflect repository architecture and framework conventions rather than underlying language difficulty.

\subsection{Oracle Regression for Qwen3 Coder and Gemini 2.5 Flash}
\label{app:appendix_oracle_regression}

Qwen3-Coder drops from 3.33\% to 2.57\%, and Gemini 2.5 Flash drops from 3.24\% to 1.98\%. Both models resolve the exact same 13 instances under Automated retrieval. This alignment is a direct consequence of embedding-based retrieval being model-independent: both models are provided with identical localized code snippets, and thus the resolved set is determined by retrieval coverage rather than model-specific repair capability. 

In contrast, tracing the instances each model fails to resolve under the Oracle setting reveals two distinct failure mechanisms.

\vspace{5pt}
\noindent\textbf{Incomplete Oracle context for new-file creations.}
The Oracle setting provides models with the set of files modified in the developer fix. However, when a fix requires creating a new file that does not exist at the base commit, that file cannot be included in the oracle context. One of the 10 instances that Qwen3-Coder fails to retain under Oracle retrieval falls into this category: \texttt{zulip-flutter-1340} requires creating \texttt{lib/model/actions.dart}, which is absent from the provided context. As a result, the model is unable to reproduce the intended fix within the constrained file set. In contrast, under Automated retrieval, the model successfully resolves this instance by patching a subset of existing files sufficient to pass the test suite, effectively bypassing the need for file creation. This option is unavailable in the Oracle setting, where the fixed file list implicitly constrains the repair scope. This failure mode does not affect Gemini 2.5 Flash, whose 10 Oracle regression instances all involve modification-only patches.

\vspace{5pt}
\noindent\textbf{Over-editing under full-file context.}
The remaining 9 of Qwen3’s and all 10 of Gemini’s regression instances had a complete oracle file list, yet both models still failed. In the Automated retrieval setting, models are provided with localized code snippets (i.e., at the function or class granularity), which constrains the search space and encourages targeted, minimal edits. In contrast, the Oracle setting exposes the model to the full file content. While this provides complete contextual information, it also increases the likelihood of modifying unrelated regions of the code. We observe that Qwen3-Coder and Gemini 2.5 Flash consistently introduce additional changes beyond the ground-truth fix, leading to regressions. This suggests that excessive context can dilute the model’s focus, resulting in over-editing behavior, whereas constrained retrieval acts as an implicit regularization mechanism that promotes more precise and semantics-preserving patches.

Table~\ref{tab:model-patch-stats} quantifies this effect across all models and settings.
Under automated retrieval, all models produce compact patches (8.5--12.7 total changed lines, $\approx$1.0 files modified), reflecting the narrowly targeted edits that result from working with localized code snippets.
Under oracle retrieval, patch sizes increase substantially (37.6--115.5 total lines, 1.3--4.9 files modified).
Gemini 2.5 Flash shows the largest expansion (8.7$\to$115.5 lines; 1.0$\to$4.9 files), followed by Claude Sonnet 4.5 (8.5$\to$71.1 lines), Qwen3 Coder (12.7$\to$57.1 lines), and GPT-5.2 (10.2$\to$37.6 lines). Notably, Claude Sonnet 4.5 expands substantially yet does not regress under oracle, suggesting that patch size alone does not determine whether over-editing introduces regressions; the pattern is most pronounced for Gemini 2.5 Flash and Qwen3 Coder, consistent with their oracle resolution rate drops.


\begin{table}[t]
\centering
\small
\setlength{\tabcolsep}{5pt}
\vspace{10pt}
\caption{Average fix patch statistics for model predictions under Automated and Oracle retrieval settings. \#L: total changed lines (added + removed). \#A: added lines. \#R: removed lines. \#H: hunks. \#F: files modified. Averages are over non-empty submitted patches only.}
\label{tab:model-patch-stats}
\begin{tabular}{lrrrrr}
\toprule
\textbf{Model} & \textbf{\#L} & \textbf{\#A} & \textbf{\#R} & \textbf{\#H} & \textbf{\#F} \\
\midrule
\multicolumn{6}{l}{\textit{Automated Retrieval}} \\
Claude Sonnet 4.5 &  8.5 &  6.2 & 2.3 & 1.4 & 1.0 \\
Gemini 2.5 Flash  &  8.7 &  5.9 & 2.8 & 1.4 & 1.0 \\
GPT-5.2           & 10.2 &  7.3 & 2.9 & 1.8 & 1.1 \\
Qwen3 Coder       & 12.7 &  9.0 & 3.7 & 1.6 & 1.1 \\
\midrule
\multicolumn{6}{l}{\textit{Oracle Retrieval}} \\
Claude Sonnet 4.5 &  71.1 & 44.6 & 26.4 & 6.9 & 3.6 \\
Gemini 2.5 Flash  & 115.5 & 93.9 & 21.6 & 9.5 & 4.9 \\
GPT-5.2           &  37.6 & 26.4 & 11.2 & 2.6 & 1.3 \\
Qwen3 Coder       &  57.1 & 41.7 & 15.4 & 4.8 & 2.3 \\
\bottomrule
\end{tabular}
\end{table}

Listing~\ref{lst:thunderbird-oracle} illustrates this for \texttt{thunderbird-android-6360} (Qwen3-Coder Oracle regression). The correct fix consists of a single insertion: adding \texttt{logBuffer.clear()} at the top of \texttt{readHelloResponse()}. The Automated retrieval patch applies exactly this change (one hunk, +1 line) and passes the test suite. In contrast, the Oracle retrieval patch includes the same insertion but, having access to the full \texttt{SmtpResponseParser.kt} file, additionally removes \texttt{logBuffer.clear()} from \texttt{readResponse()} and relocates it to \texttt{readResponseAfterReplyCode()}. This modification introduces a behavioral change across methods (three hunks, +1/$-$1 line) and leads to test failures.

Listing~\ref{lst:thunderbird-9640-oracle} shows a similar pattern in \texttt{thunderbird-android-9640} (observed for both models). The ground-truth fix introduces an early-return guard before any dependency injection occurs. The Oracle patch applies the same guard but places it one line later, after \texttt{DI.get<Clock>()} has already been invoked, which changes execution order relative to the intended fix and results in failure.

\begin{figure}[h]
\centering
\begin{minipage}[t]{0.46\linewidth}
\small\textbf{Gold / Automated} \hfill \textit{1 hunk, +1 line}\\[2pt]
\begin{Verbatim}[commandchars=\\\{\}]
\DH{@@ -33,6 +33,8 @@}
\DC{ fun readHelloResponse() \{}
\DA{    logBuffer.clear()}
\DC{     val replyCode = readReplyCode()}
\end{Verbatim}
\end{minipage}
\hfill
\begin{minipage}[t]{0.50\linewidth}
\small\textbf{Oracle} \hfill \textit{3 hunks, +1/$-$1 lines}\\[2pt]
\begin{Verbatim}[commandchars=\\\{\}]
\DH{@@ -33,6 +33,7 @@}
\DC{ fun readHelloResponse() \{}
\DA{    logBuffer.clear()}
\DC{     val replyCode = readReplyCode()}
\DH{@@ -153,7 +154,6 @@}
\DC{ fun readResponse(...) \{}
\DD{    logBuffer.clear()}
\DC{     val replyCode = readReplyCode()}
\DH{@@ -162,6 +162,7 @@}
\DC{ private fun readResponseAfterReplyCode() \{}
\DC{     logBuffer.clear()}
\end{Verbatim}
\end{minipage}
\caption{Patches for \texttt{thunderbird-android-6360} (\texttt{SmtpResponseParser.kt}). Gold inserts \texttt{logBuffer.clear()} in \texttt{readHelloResponse()}; Automated setting replicates this minimal change. Oracle makes the same correct insertion but, having seen the full file, also removes \texttt{logBuffer.clear()} from \texttt{readResponse()} and repositions it in \texttt{readResponseAfterReplyCode()}, altering runtime behavior.}
\label{lst:thunderbird-oracle}
\end{figure}

\begin{figure}[h]
\centering
\begin{minipage}[t]{0.46\linewidth}
\small\textbf{Gold / Automated} \hfill \textit{guard before DI call}\\[2pt]
\begin{Verbatim}[commandchars=\\\{\}]
\DH{@@ -85,6 +85,10 @@}
\DC{     get() \{}
\DA{        if (!isQuietTimeEnabled)}
\DA{            return false}
\DC{         val clock = DI.get<Clock>()}
\DC{         val quietTimeChecker = ...}
\DC{         return quietTimeChecker.isQuietTime}
\end{Verbatim}
\end{minipage}
\hfill
\begin{minipage}[t]{0.46\linewidth}
\small\textbf{Oracle} \hfill \textit{guard after DI call}\\[2pt]
\begin{Verbatim}[commandchars=\\\{\}]
\DH{@@ -86,12 +86,14 @@}
\DC{     get() \{}
\DC{         val clock = DI.get<Clock>()}
\DA{        if (!isQuietTimeEnabled)}
\DA{            return false}
\DC{         val quietTimeChecker = ...}
\DC{         return quietTimeChecker.isQuietTime}
\end{Verbatim}
\end{minipage}
\caption{Patches for \texttt{thunderbird-android-9640} (\texttt{K9NotificationStrategy.kt}, repeated identically in \texttt{NotificationHelper.kt}). Gold and Automated setting prepend the early-return guard before \texttt{DI.get<Clock>()}, short-circuiting the dependency injection call. Oracle inserts the guard one statement too late: the DI call executes unconditionally, failing tests that assert no DI interaction when quiet time is disabled.}
\label{lst:thunderbird-9640-oracle}
\end{figure}

Across the 7 instances lost by both models under complete oracle context, the Oracle setting patches are consistently larger than the corresponding Automated retrieval patches despite addressing the same underlying tasks. In these cases, the Automated patches pass while the Oracle patches fail. This pattern suggests that full-file context can lead Qwen3-Coder and Gemini 2.5 Flash to incorporate additional, non-essential modifications beyond the minimal fix, resulting in behaviorally different patches that no longer preserve correctness.

In contrast, Claude Sonnet 4.5 and GPT-5.2 do not exhibit this effect to the same extent, which is consistent with their positive Oracle delta.

\subsection{Fault Localization Performance}
\label{subsec:fault_localization_performance}

Table~\ref{tab:retrieval-metrics} shows that fault localization is the primary bottleneck. Overall retrieval performance ranges from 13.8--18.6\% recall, 42.9--57.2\% precision, and 17.6--23.3\% F1, with Claude Sonnet 4.5 achieving the highest scores. The large precision-recall gap (~3x across all models) indicates conservative file selection: models identify relevant files but miss many necessary modifications.


\begin{table*}[t]
\centering
\small
\setlength{\tabcolsep}{4pt}
\caption{File-level retrieval metrics (Recall / Precision / F1, macro-averaged, \%) by ground-truth file count under Automated and Oracle retrieval settings (407 tasks).}
\label{tab:retrieval-metrics}
\begin{tabular}{l r | rrr rrr rrr rrr}
\toprule
& & \multicolumn{3}{c}{\textbf{Claude~Sonnet 4.5}} & \multicolumn{3}{c}{\textbf{Qwen3-Coder}} & \multicolumn{3}{c}{\textbf{GPT~5.2}} & \multicolumn{3}{c}{\textbf{Gemini~2.5 Flash}} \\
\cmidrule(lr){3-5}\cmidrule(lr){6-8}\cmidrule(lr){9-11}\cmidrule(lr){12-14}
\textbf{\#Files} & \textbf{$n$} & R & P & F1 & R & P & F1 & R & P & F1 & R & P & F1 \\
\midrule
\multicolumn{14}{l}{\textit{Automated}} \\
1 file          &  71 & 54.9 & 54.9 & 54.9 & 46.5 & 46.5 & 46.5 & 35.2 & 35.2 & 35.2 & 35.2 & 35.2 & 35.2 \\
2--3 files      &  89 & 22.7 & 49.4 & 30.5 & 23.6 & 48.3 & 30.9 & 20.4 & 44.4 & 27.5 & 20.2 & 44.9 & 27.5 \\
4--5 files      &  52 & 12.1 & 51.9 & 19.6 & 12.7 & 53.8 & 20.5 &  9.9 & 40.4 & 15.8 & 10.5 & 42.3 & 16.7 \\
6--10 files     &  70 &  9.4 & 65.7 & 16.4 &  7.5 & 52.1 & 13.0 &  7.4 & 49.3 & 12.8 &  6.7 & 46.4 & 11.6 \\
11+ files       & 125 &  2.9 & 61.6 &  5.5 &  3.0 & 53.6 &  5.5 &  2.1 & 43.6 &  3.9 &  2.3 & 44.0 &  4.3 \\
\cmidrule(lr){1-14}
Overall         & 407 & 18.6 & 57.2 & 23.3 & 17.1 & 51.0 & 21.4 & 13.8 & 42.9 & 17.6 & 13.8 & 42.9 & 17.6 \\
\midrule
\multicolumn{14}{l}{\textit{Oracle}} \\
1 file          &  71 & 95.8 & 95.8 & 95.8 & 97.2 & 96.5 & 96.7 & 84.5 & 83.8 & 84.0 & 100.0 & 100.0 & 100.0 \\
2--3 files      &  89 & 57.5 & 94.9 & 68.9 & 50.2 & 98.5 & 64.4 & 40.6 & 89.9 & 54.9 &  60.9 &  96.5 &  71.6 \\
4--5 files      &  52 & 48.8 & 89.8 & 58.1 & 35.9 & 93.2 & 48.0 & 21.7 & 89.2 & 34.4 &  50.2 &  92.6 &  60.7 \\
6--10 files     &  70 & 43.1 & 90.8 & 53.6 & 30.4 & 89.9 & 41.5 & 14.2 & 87.1 & 23.2 &  43.2 &  95.8 &  53.6 \\
11+ files       & 125 & 22.8 & 80.2 & 31.2 & 13.6 & 77.1 & 20.0 &  8.3 & 75.6 & 12.0 &  35.1 &  88.9 &  43.0 \\
\cmidrule(lr){1-14}
Overall         & 407 & 50.0 & 89.2 & 58.0 & 41.9 & 89.4 & 50.4 & 31.4 & 83.9 & 38.7 &  55.4 &  94.2 &  63.3 \\
\bottomrule
\end{tabular}
\end{table*}

\vspace{5pt}
\noindent\textbf{Recall Degradation by File Count.}
Recall degrades severely with file count: from 35--55\% on single-file tasks to 2--3\% on 11+ file tasks, representing a 33--53 percentage point decline. For tasks requiring 6--10 files, models identify fewer than 10\% of necessary files. This catastrophic degradation reflects models' inability to trace cross-layer dependencies in mobile architectures, where changes cascade across network, data, presentation, view, and test layers.

\begin{figure*}[t]
\centering
\includegraphics[width=\textwidth]
{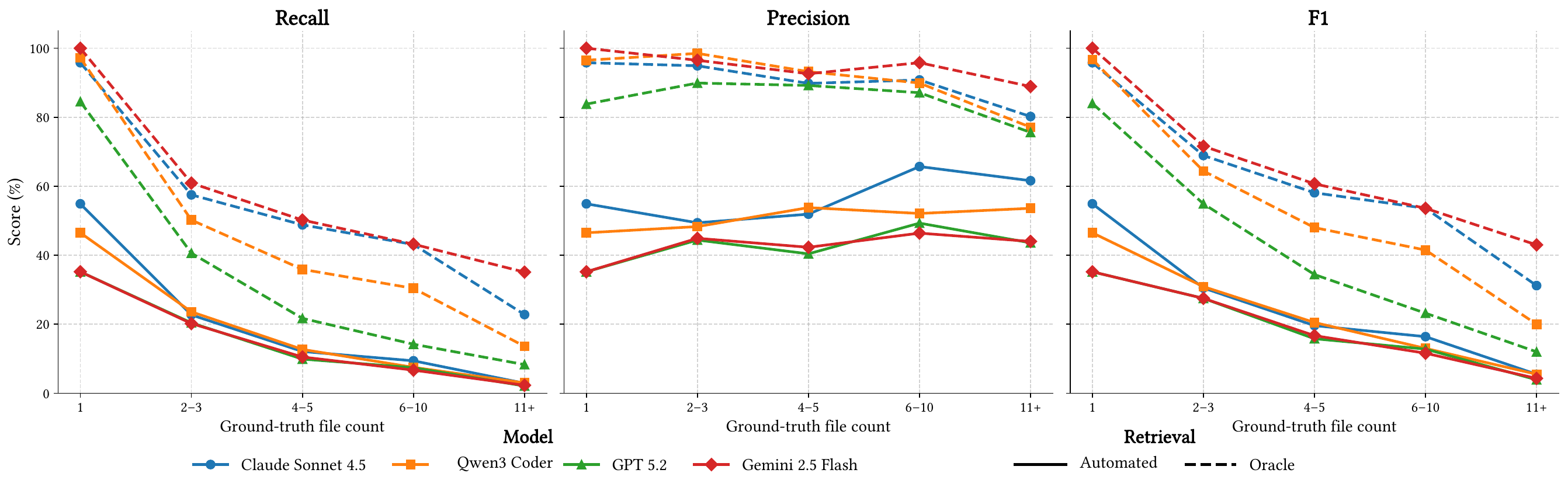}
\caption{Retrieval metrics (recall, precision, F1) by ground-truth file count across all four models. Recall collapses from 35--55\% on single-file tasks to 2--3\% on 11+ file tasks, while precision remains comparatively stable (35--66\%), widening the precision-recall gap with complexity. F1 degrades sharply, driven entirely by recall failure.}
\label{fig:retrieval-metrics-filecount}
\end{figure*}

\vspace{5pt}
\noindent\textbf{Precision Trends.}
Precision increases with file count (35--55\% for single-file tasks to 46--66\% for 6--10 file tasks), indicating conservative behavior on complex changes. Localization errors stem primarily from missed files rather than false positives: models identify a core subset correctly but fail to expand to all necessary modifications.

\vspace{5pt}
\noindent\textbf{Localization Patterns.}
Three distinct patterns emerge:

\begin{itemize}
    \item \textbf{Perfect localization} (6--10\% of tasks): Almost exclusively single-file changes with strong semantic signals in issue descriptions.
    \item \textbf{Zero localization} (43--57\% of tasks): Models edit entirely incorrect files, common when ground truth modifies non-code artifacts or peripheral modules. 111 tasks (27\%) show zero recall across all four models simultaneously.
    \item \textbf{Partial localization} (37--47\% of tasks): Models find some relevant files but miss cascading modifications across architectural layers.
\end{itemize}

\subsubsection{Implications}
Both file-level localization and patch generation are substantial bottlenecks.
Under automated retrieval, recall (13.8\%–18.6\%) sets a hard ceiling on resolution: actual resolution (3.23\%–4.23\%) falls below this ceiling, so localization is one binding constraint. Under oracle retrieval, recall rises to 31.4\%–55.4\% overall and 84.5\%–100\% on single-file tasks, yet resolution still tops out at 5.69\%. Generation is therefore the larger remaining
constraint: improving localization to perfect lifts resolution from 4.23\% to 5.69\% (a +1.46pp gain), whereas closing the oracle generation gap on already-localized cases offers substantially more headroom.

Improving repair performance, therefore, requires advances on both axes. Prior work on automated repair has primarily focused on patch generation~\cite{monperrus2018automatic,ding2020patching}; our results suggest that fault localization in mobile codebases warrants comparable attention. Promising directions for localization include dependency-aware retrieval to capture
inter-file relationships, incorporation of architectural priors (that is, leveraging knowledge of common software design patterns and component interactions such as MVVM or repository-based data flow), and specialized retrieval strategies for non-code artifacts that are underrepresented in embedding spaces. For generation, the persistent oracle gap points to multi-file consistency constraints, framework-aware reasoning about lifecycle and resource semantics, and explicit modeling of cross-artifact dependencies between source code, resources, and build configurations.

\section{Discussion}
\label{sec:discussion}

Our evaluation reveals overall resolution rates of 3.23\%–4.23\% under automated retrieval and at most 5.69\% under oracle retrieval on {\benchmark}. Three architectural properties of mobile development explain this low performance. First, multi-artifact coordination: single-artifact tasks comprise 59\% of the benchmark and account for 96.4\% of resolved instances at per-task resolution rates of 5.0\%–6.7\%, while multi-artifact tasks comprise 41\% of the benchmark and account for only 3.6\% of resolved instances at 0.0\%–0.6\%. 
Second, framework-mediated execution: platform lifecycles obscure control flow from static analysis.
Third, cross-module dependencies: recall drops from 35.2\%–54.9\% on
single-file tasks to 6.7\%–9.4\% on tasks requiring 6–10 files, and further to 2.1\%–3.0\% on tasks with 11 or more files.

Both file-level localization and patch generation are substantial
bottlenecks. Under automated retrieval, recall (13.8
achievable resolution at the recall ceiling. Under oracle retrieval, recall rises to 31.4\%–55.4\% overall and 84.5\%–100\% on single-file tasks, yet resolution still reaches at most 5.69\%, indicating that patch generation is the larger remaining bottleneck. Progress on mobile app issue resolution therefore requires advances on both axes: multi-file localization (dependency tracing, artifact-aware retrieval, framework-specific reasoning) and patch generation under multi-file and multi-artifact coordination constraints.

Resolution rates decline monotonically with file count: single-file tasks achieve 12.7\%–15.5\% resolution, 2–3 file tasks drop to 2.2\%–4.5\%,
mid-complexity tasks (4–10 files) fall to 0.0\%–3.8\%, and tasks with 11 or more files yield 0\% across all models. This pattern suggests models struggle with architectural complexity, that is, interconnected changes across heterogeneous components, rather than raw file count alone.




\section{Future Work}
Our findings suggest three high-priority research directions. First, given that localization is the primary bottleneck, techniques combining semantic search with static analysis for multi-file dependency tracing warrant immediate investigation. Second, repository-specific success patterns hint that certain architectural characteristics aid automated repair; systematically analyzing what makes codebases "LLM-friendly" could inform both tool design and best practices. Third, mobile issues often include screenshots and crash logs; evaluating multi-modal models' ability to ground repairs in visual artifacts may improve UI-related bug localization. Extending to iOS, developing framework-aware representations encoding lifecycle semantics, and testing iterative agentic workflows provide additional promising directions.

\section{Limitations}
\label{sec:limitations}
{\benchmark} focuses on Android Native and cross-platform apps, including React-Native and Flutter repositories, and does not include iOS-native projects written in Swift or Objective-C.
This exclusion is a deliberate reproducibility constraint rather than a coverage gap. iOS builds are tightly coupled to Xcode and macOS: the Xcode and Apple SDKs Agreement ~\citep{Apple2024XcodeAgreement} explicitly prohibits execution on non-Apple-branded hardware, and macOS itself cannot be legally run in standard Linux container environments. 
Consequently, the containerized build-and-test execution that is central to our evaluation design is not achievable for iOS repositories without proprietary infrastructure. 
Extending MobileDev-Bench to iOS would require macOS-based CI runners or hardware-in-the-loop setups, both of which introduce environment heterogeneity that undermines the reproducibility guarantees the benchmark is designed to provide.
Evaluation relies on execution-based validation using repository test suites, which may not fully capture issues with insufficient or incomplete test coverage. Additionally, the benchmark is constructed from issue–pull request pairs in open-source repositories, which may not represent all mobile development activities, such as UI design iterations or refactoring tasks without associated issues. Finally, our evaluation uses a single-pass repair pipeline, whereas app development often involves iterative debugging and interactive workflows.


\section{Potential Societal Impact}
\label{sec:potential-impact}

MobileDev-Bench provides an execution-validated benchmark for mobile-specific program repair, exposing failure modes invisible to library-centric evaluations, guiding practitioners on where LLM assistants are and are not reliable in mobile CI pipelines, and highlighting the need for human oversight when deploying model-generated patches in accessibility-, localisation-, or security-sensitive components.

\noindent\textbf{AI Assistance.} AI-based writing tools were used only for minor grammatical and wording improvements. All technical content and analyses were produced and verified by the authors.


\end{document}